\documentclass[jcp,12pt,reprint,onecolumn]{revtex4-1}%
\usepackage{amsmath,amsfonts,amssymb,amsbsy,eucal,dsfont}
\usepackage{natbib}
\usepackage{color}
\usepackage[colorlinks]{hyperref}
\usepackage{rotating}
\usepackage{graphicx}%
\usepackage{amssymb}
\ifx\pdfoutput\relax\let\pdfoutput=\undefined\fi
\newcount\msipdfoutput
\ifx\pdfoutput\undefined\else
\ifcase\pdfoutput\else
\ifx\paperwidth\undefined\else
\ifdim\paperheight=0pt\relax\else\pdfpageheight\paperheight\fi
\ifdim\paperwidth=0pt\relax\else\pdfpagewidth\paperwidth\fi
\fi\fi\fi
\begin{document}

\preprint{HEP/123-qed}

\title{Rotational self-diffusion in suspensions of charged particles: Revised Beenakker-Mazur and Pairwise Additivity methods versus numerical simulations}

\author{K. Makuch}
\affiliation{Faculty of Physics, Institute of Theoretical Physics, University of Warsaw, ul. Pasteura 5, 02-093 Warsaw, Poland}
\email{Karol.Makuch@fuw.edu.pl}
\author{M. Heinen}
\affiliation{Division of Chemistry and Chemical Engineering, California Institute of Technology, Pasadena, California 91125, USA}
\author{G. C. Abade}
\affiliation{ Fachbereich Physik, Universit\"at Konstanz, 78457 Konstanz, Germany
\footnote{On the leave of absence from: Departamento de Engenharia
 Mec\^anica, Universidade de Bras\'{\i}lia, Campus Darcy Ribeiro,
70910-900, Asa Norte, Bras\'{\i}lia-DF, Brazil}}              
\author{G. N\"{a}gele}
\affiliation{Institute of Complex Systems (ICS-3), Research Centre J\"{u}lich, 52425 J\"{u}lich, Germany}

\pacs{PACS number}

\begin{abstract}

To the present day, the Beenakker-Mazur (BM) method is the most comprehensive statistical physics approach to the calculation of short-time transport properties of colloidal suspensions. A revised version of the BM method with an improved treatment of hydrodynamic interactions is presented and evaluated regarding the rotational short-time self-diffusion coefficient, $D^r$, of suspensions of charged particles interacting by a hard-sphere plus screened Coulomb (Yukawa) pair potential. To assess the accuracy of the method, elaborate simulations of $D^r$ have been performed,  covering a broad range of interaction parameters and particle concentrations. The revised BM method is compared in addition with results by a simplifying pairwise additivity (PA) method in which the hydrodynamic interactions are treated on a two-body level. The static pair correlation functions required as input to both theoretical methods are calculated using the Rogers-Young integral equation scheme. While the revised BM method reproduces the general trends of the simulation results, it systematically and significantly underestimates the rotational diffusion coefficient. The PA method agrees well with the simulation data at lower volume fractions, but at higher concentrations $D^r$ is likewise underestimated.  For a fixed value of the pair potential at mean particle distance comparable to the thermal energy, $D^r$ increases strongly with increasing Yukawa potential screening parameter. 
\end{abstract}

\startpage{1}
\endpage{ }
\maketitle

\section{Introduction}

Short-time transport properties of colloidal suspensions such as translational self- and collective diffusion coefficients, hydrodynamic function and high-frequency viscosity have been the subject of numerous experimental studies
\cite{VanMegen1985,Megen_552_91_1989,PhysRevE.52.2707,PhysRevLett.84.785,fritz2002high,orsi2012dynamics,westermeier2012structure}. These studies have been accompanied over the past years by computer simulation studies (see, e.g. \cite{JChemPhys_93_3484,heinen2011short}) 
and theoretical works 
\cite{einstein1906neue,saito1950concentration,batchelor1972determination, resummation83,Beenakker1984effective,Beenakker1984Diffusion,nozieres1987local,cichocki1989effective, felderhof1990hydrodynamics,trojczastkowalepkosc}. While short-time transport properties are expressible theoretically as rather simple equilibrium averages invoking hydrodynamic mobility tensors, the difficulty in their actual calculation arises from the long-ranged many-body hydrodynamic interactions (HIs) between the particles. The slowing influence of the HIs is particularly pronounced when particles are in relative motion close to each other.

There exist two major solution schemes which have been used for the calculation of short-time properties. The first one is the so-called pairwise additivity (PA) approximation. In its most complete version, all two-body HIs contributions are accounted for including lubrication terms, but three-body and higher order interaction contributions are disregarded. By construction, its application range is usually limited to semi-dilute systems, and here in particular to charge-stabilized suspensions where near-contact configurations of particles are statistically unlikely \cite{heinen2011short}. Regarding short-time self-diffusion coefficients and the high-frequency viscosity, however, the PA method can be profitably used also at higher concentrations, owing to the steep decay, with increasing inter-particle distances, of the hydrodynamic mobility tensors associated with these quantities.

Different from the PA scheme, the semi-analytical method of calculating short-time diffusion and viscosity properties by Beenakker-Mazur \cite{resummation83,Beenakker1984effective,Beenakker1984Diffusion}, commonly referred to as the 
$\delta\gamma$ method, is applicable also to concentrated suspensions. It is a mean-field-type approximation method which accounts for many-body HIs contributions in the form of so-called ring self-correlation diagrams, 
without an account of lubrication effects. In its standard second-order version, the only required external input is the static structure factor $S(q)$ as function of the scattering wavenumber $q$. A major short-coming of the original $\delta\gamma$ method is its poor treatment of the translational short-time self-diffusion coefficient $D^{t}$. This deficiency can be overcome by using a more accurate method for the self-part, e.g. by 
using the PA approximation result for $D^t$ at lower concentrated systems. 
The self-part corrected $\delta\gamma$ scheme has been applied both to suspensions of neutral and charge-stabilized particles \cite{genz1991collective,heinen2011short,banchio2008short}, for the calculation of the  hydrodynamic and collective diffusion functions, and the high-frequency viscosity. It was applied recently also to suspensions of hydrodynamically structured particles \cite{riest2014submitted}, and with additional approximations also to a binary hard-sphere mixture \cite{wang2014short}.
The predictions for these systems are decently good, with inaccuracies revealed at all concentrations. These inaccuracies can be partially attributed to the approximate treatment of the HIs in the $\delta\gamma$ method, and partially to the invoked mean-field approximation. In recent work by Makuch and Cichocki \cite{makuch2012transport}, the approximation steps in the original derivation of the $\delta\gamma$ method have been reduced, in particular by accounting for a large number of hydrodynamic multipoles in the truncated multipolar matrices which are extrapolated to infinite order. The observation that the revised $\delta\gamma$ method by Makuch  and Cichocki, with its improved hydrodynamic mobility tensor treatment, has not resulted in a systematic improvement of the hydrodynamic function and high-frequency viscosity predictions points to a fortuitous cancellation of errors in the various approximation steps of the original method by Beenakker and Mazur.

Experimental studies of the short-time rotational self-diffusion coefficient, $D^r$, in concentrated suspensions are based on techniques which can distinguish different particle orientations. These methods include depolarized dynamic light scattering on optically anisotropic particles \cite{PhysRevE.52.2707}, nuclear magnetic resonance \cite{kanetakis1997rotational}, time-resolved phosphorescence anisotropy \cite{koenderink2001rotational, koenderink2003validity, PhD_Koenderink} and polarized fluorescence recovery after photobleaching \cite{lettinga2004rotational}. Most experimental work 
on rotational diffusion has been for monodisperse colloidal systems of neutral hard spheres and charge-stabilized particles. In addition, binary mixtures of charge-stabilized particles have been studied where one component is very dilute \cite{koenderink2001rotational, koenderink2003validity, PhD_Koenderink}. 

Simulation work on rotational diffusion has dealt so far mainly with monodisperse systems of non-permeable \cite{hagen1999rotational, phillips1988hydrodynamic, banchio2008short} and permeable hard spheres \cite{abade2011rotational}. Charge-stabilized systems have been addressed in few simulation studies only \cite{hagen1999rotational, banchio2008short}, focused on low-salinity systems with weak electrostatic screening. 
Therefore, a systematic simulation study of rotational diffusion in charge-stabilized systems is still on demand.

Short-time rotational self-diffusion in hard-sphere suspensions was studied theoretically by various groups using truncated hydrodynamic cluster expansions up to quadratic order in the particle volume fraction $\phi$ \cite{urdaneta1989calculation, JCP_96_3137, clercx1991retarded, PhysRevE.52.2707}. The high-precision result by Cichocki {\em et al.} \cite{trojczastkowasamodyfuzja},
\begin{equation} \label{eq:virial-HS-second}
  \frac{D^r}{D_0^r} = 1 - 0.631\phi - 0.726\phi^2 +{\cal O}(\phi^3)\,,
\end{equation}
includes a lubrication correction for the three-body HIs contributions. Here, $D_0^r=k_B T/(8\pi\eta a^3)$ is the single-particle rotational diffusion coefficient of a no-slip sphere of hydrodynamic radius $a$. As shown in \cite{banchio2008short, abade2011rotational}, 
Eq. (\ref{eq:virial-HS-second}) describes simulation and experimental data remarkably well for volume fractions up to the freezing transition value $\phi_f = 0.494$, indicating that higher-order virial coefficients are small or mutually cancel out. 

Regarding rotational diffusion, the $\delta\gamma$ method has been applied so far to hard-sphere suspensions only, in the work by Treloar and Masters \cite{treloar1989short} where approximations along the line of those introduced by Beenakker and Mazur have been made. It was not applied so far to rotational diffusion in charge-stabilized suspensions. However, these systems have been analyzed in the weak electrostatic screening regime using a simplified PA approximation approach based on a truncated inverse distance expansion of the two-body rotational mobility tensors, by considering in addition the leading-order long-distance hydrodynamic three-body term \cite{zhang2002tracer, koenderink2003validity, Watzlawek1997}. For charged particles with strong long-distance repulsion, the remarkable scaling relation, 
\begin{equation} \label{eq:CSdeionized}
  \frac{D^r}{D_0^r} = 1 - a_r\;\!\phi^2\,,
\end{equation}
with $a_r \approx 1.3$ has been obtained by this simplifying approach. The comparison with Lattice-Boltzmann \cite{hagen1999rotational} and accelerated Stokesian dynamics simulation results \cite{banchio2008short} has shown that it applies accurately up to $\phi \approx 0.3$. Different from Eq. (\ref{eq:virial-HS-second}) valid for neutral hard spheres, the scaling relation in Eq. (\ref{eq:CSdeionized}) is not a second-order virial expansion result. It originates basically from the $\phi^{-1/3}$ concentration scaling of the radius, $r_m$, of next-neighbor shells in low-salinity systems \cite{banchio2008short}. 

The present article includes a comprehensive theoretical and simulation analysis of short-time rotational diffusion in fluid-like, charged particles suspensions whose static pair interactions are modeled by the hard-sphere plus repulsive Yukawa (HSY) pair potential. A revised version of the $\delta\gamma$ method for $D^r$ by Treloar and Masters 
is evaluated for a broad range of volume fractions and reduced pair potential strengths, and two screening parameters characteristic for the weak and strong screening regimes, respectively. The accuracy of the revised $\delta\gamma$ method for $D^r$ is assessed in the comparison with high-accuracy simulation results which we have obtained using a multipole simulation method encoded in the HYDROMULTPOLE simulation package \cite{trojczastkowasamodyfuzja}. The revised $\delta\gamma$ method and simulation results are compared in addition with our predictions by a simplifying pairwise additivity (PA) method in which the rotational hydrodynamic mobility tensors are treated on the two-body level, without an additional long-distance truncation as made in earlier work. The radial distribution function (RDF), $g(r)$, and static structure factor of the HSY model constituting the static input to the theoretical methods are calculated using the accurate Rogers-Young (RY) integral equation scheme. The present work is the first systematic theory-simulation study of rotational self-diffusion in the HSY model.   

The article is organized as follows: In Sec. \ref{sec:Dr_short}, we give the essentials of short-time rotational self-diffusion. Sec. \ref{sec:static-correlations} includes the description of the  HSY model with employed interaction parameters, and a discussion of the RY radial distribution functions used in the analytical-theoretical calculation of $D^r$. The revised Beenakker-Mazur method of calculating $D^r$ is explained in Sec. \ref{sec:Revised-BM}. In Secs. \ref{sec:sumulationmethod} and \ref{sec:PA}, respectively, the employed simulation and PA methods are described. Our results are presented in Sec. \ref{sec:results}, 
and our finalizing conclusions in Sec. \ref{sec:conclusions}.           

\section{Short-time rotational self-diffusion coefficient}\label{sec:Dr_short}

We consider rotational diffusion in a fluid-state suspension of monodisperse spherical Brownian particles immersed in a structureless Newtonian solvent of shear viscosity $\eta$, on a coarse-grained Brownian time scale 
exceeding the rotational and translational momentum relaxation times $\tau_B^t \sim \tau_B^r$, respectively, 
by several orders of magnitude \cite{nagele1996dynamics,PhysRevE.52.2707}. 
On this scale, particles and fluid move quasi-inertia-free, and the fluid-mediated HIs are 
acting quasi-instantaneously. The configurational evolution of the particles is then governed by the 
generalized Smoluchowski equation 
\cite{russel1992colloidal, pusey1991liquids, Dhont1996} for the $N$-particle probability density function 
$p\left(\mathbf{R}_{1},\ldots ,\mathbf{R}_{N},\mathbf{\hat{u}}_{1},\ldots ,\mathbf{\hat{u}}_{N},t\right)$ of the 
sphere center positions $\mathbf{R}_{1},\ldots ,\mathbf{R}_{N}$ and orientations $\mathbf{\hat{u}}_{1},\ldots ,\mathbf{\hat{u}}_{N}$
at time $t$. The associated low-Reynolds-number incompressible fluid flow is described by the linear stationary Stokes equation \cite{kim1991microhydrodynamics}. The hydrodynamic ingredients to the generalized Smoluchowski 
equation derived from the Stokes equation are the translational-rotational mobility tensors quantifying  
the linear relations, 
\begin{eqnarray}
\mathbf{U}_{i} &=&\sum_{j=1}^{N}\boldsymbol{\mu }_{ij}^{tt}\left( \mathbf{R}%
_{1}\ldots \mathbf{R}_{N}\right) \cdot \mathbf{F}_{j}+\sum_{j=1}^{N}%
\boldsymbol{\mu }_{ij}^{tr}\left( \mathbf{R}_{1}\ldots \mathbf{R}_{N}\right)
\cdot \mathbf{T}_{j}, \label{eq:lin_mapping_TF_to_U} \\
\mathbf{\Omega }_{i} &=&\sum_{j=1}^{N}\boldsymbol{\mu }_{ij}^{rt}\left( 
\mathbf{R}_{1}\ldots \mathbf{R}_{N}\right) \cdot \mathbf{F}%
_{j}+\sum_{j=1}^{N}\boldsymbol{\mu }_{ij}^{rr}\left( \mathbf{R}_{1}\ldots 
\mathbf{R}_{N}\right) \cdot \mathbf{T}_{j}\,, \label{eq:lin_mapping_TF_to_Omega}
\end{eqnarray}
between the forces and torques, $\mathbf{F}_{i}$ and $\mathbf{T}_{i}$, respectively, acting on the colloidal spheres, and the resulting translational and rotational particle velocities $\mathbf{U}_{i}$ and $\mathbf{\Omega }_{i}$. For the uniformly assumed no-slip hydrodynamic surface boundary condition, the mobility tensors are independent of the particle orientations.  

In depolarized dynamic light scattering \cite{berne2000dynamic,PhysRevE.52.2707}, the short-time rotational self-diffusion coefficient of Brownian spheres is determined from the initial decay of the measurable orientational self-correlation function, for correlation times $t$ within $\tau_B^r \ll t \ll 1/D^r_0$ where particle orientations and positions have changed by very small amounts only, on the characteristic length scale of the suspension. 
For a concentrated isotropic suspension, $D^{r}$ can be computed as the ensemble average of the trace of the rotational mobility tensor \cite{jones1988rotational}, 
\begin{equation}
D^{r}=\frac{k_{B}T}{3}\lim_{\infty}\left[\text{Tr}\left\langle \frac{1}{N}%
\sum_{i=1}^{N}\boldsymbol{\mu }_{ii}^{rr}\left( \mathbf{R}_{1}\ldots \mathbf{%
R}_{N}\right) \right\rangle \right].  \label{rot self diff}
\end{equation}%
where $\left\langle\cdots\right\rangle$ is an equilibrium ensemble average, and where the thermodynamic limit $N\to \infty$ at fixed particle concentration has been taken. It should be noted that $D^r$ as given in Eq. (\ref{rot self diff}) is, for non-zero concentrations, different from the initial slope of the mean-squared displacement of the particle orientation 
unit vector $\mathbf{\hat{u}}_{i}(t)$. 

\section{Static correlations of HSY particles} \label{sec:static-correlations}

The pair potential in the 
hard-sphere plus repulsive Yukawa (HSY) model is given by
\begin{equation}
\frac{u\left( r\right) }{k_{B}T}=\left\{ 
\begin{array}{cc}
\gamma\:\!\dfrac{\exp\{-\kappa (r -\sigma)\}}{r/\sigma}\,, & r>\sigma \\ 
\infty\,,  & r<\sigma%
\end{array}%
\right. \,,  \label{eq:HSY_pot_gk}
\end{equation}%
where $\gamma\geq 0$ is the coupling parameter of the Yukawa-type potential part, $\sigma=2a$ the hard-core diameter, $r$ the center-to-center distance between two spheres, $k_B$ the Boltzmann constant, and $T$ the absolute temperature. 
The range of the HSY potential is set by the inverse of the screening parameter $\kappa \geq 0$. In the infinite screening limit $\kappa \to \infty$, or likewise for $\gamma=0$, the hard-sphere potential is recovered. In the opposite limit $\kappa \to 0$ of zero-screening, a one-component plasma-like system is described.  

Dispersions which can be described by the HSY model range from charge-stabilized suspensions of rigid colloidal spheres \cite{holmqvist2010long} to globular protein solutions \cite{heinen2012viscosity} and dusty plasmas \cite{ivlev2012complex}.  
The HSY potential form is in general a good approximation to the state-dependent effective pair-potential between charged colloidal spheres. The latter is obtained from integrating out the degrees of freedom of the microions and 
solvent molecules. In many experimentally encountered suspensions, the short-ranged van der Waals attraction neglected in the HSY model is of no relevance, either since the electrostatic repulsion 
between the particles is strong enough to prevent near-contact configurations \cite{westermeier2012structure}, or the 
solvent dielectric constant nearly matches that of the suspended
particles \cite{Philipse1988, heinen2010short, heinen2011pair, Heinen2011Erratum}, or the particles are
sterically stabilized by grafted polymers \cite{vanGruithuijsen2013}. 
The complicated dependencies of the state-dependent potential parameters $\gamma$ and $\kappa$ in charge-stabilized suspensions on the salt ion concentration, colloidal volume fraction, and bare and effective colloidal surface charges is the topic of on-going research that covers experiments, theory and computer simulations
\cite{Trizac2004, Shapran2005, Dobnikar2006, Castaneda-Priego2006, Ruiz-Reina2008, Rojas-Ochoa2008, Labbez2009,
Calero2010, Falcon-Gonzalez2010, Barr2011, Heinen2014, HeinenPalbergLoewen2014}.
The present work is not concerned with a first-principles determination of the state-dependence of $\gamma$ and $\kappa$ which is also influenced by the specific surface electrochemistry of the dispersed particles. Instead, $\gamma$ and $\kappa$ are treated quite generally as individually variable parameters. Note further that the direct interactions in the HSY model are treated as pairwise additive. Non-pairwise additivity effects in the direct interaction of charged colloidal particles are usually quite small \cite{Loewen1998}.
\begin{figure}
\begin{centering}
\fbox{\includegraphics[width=.6\textwidth]%
{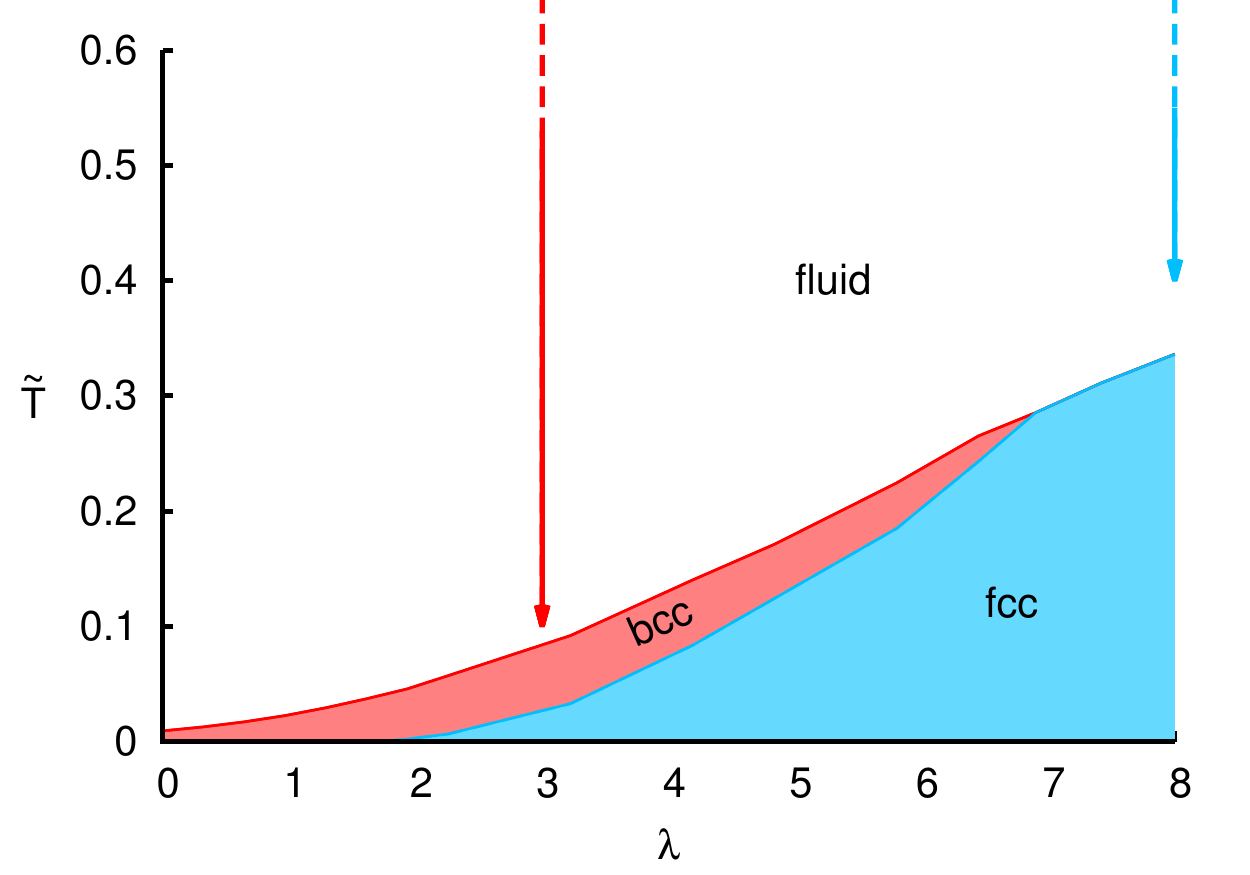}}%
\vspace{1em}
\caption{Schematic $\left(\lambda,\tilde{T}\right)$ phase diagram of the hard-core plus repulsive Yukawa (HSY) 
system with masked
hard-core interactions (point-Yukawa system). The arrows indicate the two fluid-state pathways towards the fluid-bcc and fluid-fcc phase boundary lines, respectively, followed in our calculations
of the short-time rotational self-diffusion coefficient.  
The left arrow corresponds to $\lambda=3$ and $\tilde{T}=0.1, 0.2, 0.5, 1, 2, 5, 10,
20, 50, 100, 200, 500, 1000$, and the right one to $\lambda=8$ and 
$\tilde{T}=0.4, 0.6, 0.8, 1, 2, 5, 10, 20, 50, 100, 200, 500, 1000$. The volume fractions considered in both pathways are 
$\phi=0.05, 0.15, 0.25, 0.35$.
}
\label{fig phase diagram}%
\end{centering}
\end{figure}

According to Eq.~\eqref{eq:HSY_pot_gk}, the thermodynamic state of the HSY model, and likewise the RDF as a function of $r/\sigma$, are fully characterized by three independent non-dimensional parameters which can be taken as $\kappa\sigma$, $\gamma$ and the particle volume fraction
\begin{equation}
 \phi=\pi \sigma^{3} n/6\,,
\end{equation}
 where $n$ is the number density of particles. For truly charge-stabilized suspensions, however, the physical hard-core is masked by the strong Yukawa repulsion, i.e. the likelihood for two or more spheres being in contact is negligibly small. The appropriate physical length scale for these so-called point-Yukawa systems is the geometric mean particle distance $\langle r \rangle = n^{-1/3}$, and the thermodynamic state and in particular the phase boundaries are determined by two parameters only. The phase boundaries of the point-Yukawa system look particularly simple, with nearly straight lines, in the two-dimensional $(\lambda,\tilde{T})$ phase diagram representation \cite{hamaguchi1997triple, stevens1993melting} where  
\begin{eqnarray}
\lambda &=& \kappa\;\!\langle r\rangle\,, \label{eq:lambda} \\
\tilde{T} &=&\frac{k_{B}T}{u\left(\left\langle r\right\rangle \right)}\,, \label{eq:T}
\end{eqnarray}
are the reduced screening parameter and the inverse reduced Yukawa interaction parameter, respectively. In terms of these  parameters, the dominating Yukawa-part of the HSY potential reads 
\begin{equation}
\frac{u\left(x\right)}{k_{B}T}= \frac{\exp\{-\lambda\left(x-1\right)\}}{\tilde{T}} \,,
\end{equation}
where $x=r/\langle r \rangle$ with $x > \sigma/\langle r\rangle$. If considered as a function of $x$ and $q \langle r \rangle$, the RDF and static structure factor of point-Yukawa particles are uniquely determined by the state point $(\lambda,\tilde{T})$. 

A sketch of the $(\lambda,\tilde{T})$ phase diagram of point-Yukawa particles is given in Fig.~\ref{fig phase diagram}. It consists of a high-temperature supercritical fluid phase, separated by a fluid-solid coexistence boundary from a face-centered-cubic (fcc) solid phase region at high screening, and a body-centered-cubic (bcc) solid phase region at low screening. There is a single triple point of three-phase coexistence at $\lambda_t \approx 6.9$ \cite{hamaguchi1997triple,hynninen2003phase}. According to \cite{gapinski2012freezing, gapinski2014freezing}, the fluid-solid coexistence boundary determined in simulations is well reproduced by the RY integral equation scheme in conjunction with the Hansen-Verlet criterion $S(q_m)= 3.1$ for the onset of freezing, where $q_m$ is the wavenumber position of the static structure factor maximum. 

The full phase diagram of the HSY model including systems with significantly non-zero RDF contact values $g(r=\sigma^+) >0$ is more complicated, and requires the specification of a third reduced parameter in addition to, say, $\lambda$ and $\tilde{T}$, namely the volume fraction $\phi$. As shown in simulations by Hynninnen and Dijkstra \cite{hynninen2003phase},  
there is then an additional triple point at very small $\lambda$ associated with large volume fractions 
where the fcc phase is favored. Provided the coupling parameter in Eq. (\ref{eq:HSY_pot_gk}) is sufficiently large (i.e. $\gamma \geq 20$) and $\phi$ sufficiently small (i.e. $\phi<0.5$), the phase coexistence lines of the HSY model can be essentially mapped to those of the point-Yukawa model, by expressing $\gamma$ and $\kappa\sigma$ in Eq. (\ref{eq:HSY_pot_gk}) in terms of $\{\lambda,\tilde{T},\phi\}$ using $\langle r \rangle \propto \phi^{-1/3}$. Note that for $\tilde{T}>1$, the potential energy of the Yukawa potential part at mean particle distance is smaller than the thermal energy $k_B T$. With increasing $\tilde{T}$ and fixed $\lambda$ and $\phi$, the importance of the Yukawa potential part diminishes, and the suspension becomes increasingly hard-sphere like.   
 
The present study of short-time rotational diffusion is restricted to the fluid phase regime. However, it is interesting to compare changes in rotational diffusion when the fluid-bcc and fluid-fcc parts of the fluid-solid coexistence lines are approached, respectively, on decreasing the reduced inverse Yukawa interaction parameter $\tilde{T}$. To this end, in our simulation and theoretical calculations of $D^r$ we follow two distinct pathways indicated by the two arrows in the $\left(\lambda,\tilde{T}\right)$ diagram in Fig.~\ref{fig phase diagram}. The left pathway is the vertical line along $\lambda=3$ with the reduced inverse Yukawa interaction parameter series $\tilde{T} \in \{0.1, 0.2, 0.5, 1, 2, 5, 10,
20, 50, 100, 200, 500, 1000\}$, where the smallest value $\tilde{T}=0.1$ describes a state point close to the fluid-bcc phase boundary line part of the point-Yukawa phase diagram. The right pathway in the figure is the line along $\lambda=8$
with values $\tilde{T} \in \{0.4, 0.6, 0.8, 1, 2, 5, 10, 20, 50, 100, 200, 500, 1000\}$. Here, 
the lowest value $\tilde{T}=0.4$ is close to the fluid-fcc phase boundary line part. 
For both pathways, the volume fraction is selected as $\phi=0.05$, $0.15$, 
$0.25$ and $0.35$, respectively. This amounts to simulation-based calculations of $D^r$ at $104$ different fluid-phase state points.   
The static structure factors $S(q)$ and the associated RDFs $g(r)$ of all HSY systems explored in the present work
have been calculated using the RY integral equation scheme described in 
the following subsection. We have checked that each of the considered $S(q)$'s qualifies as a liquid-state 
structure factor according to the empirical Hansen-Verlet criterion. This criterion states that at freezing into a solid  phase, $S(q_m)$ attains a value near to $3.1$ for point-Yukawa systems. For HSY systems with RDF contact values $g(\sigma^+)$ significantly larger than zero, the values of $S(q_m)$ at freezing vary in 
between $3.1$ and $2.85$, with the lower value attained by a pure hard-sphere system \cite{Hansen1969, Kremer1986, stevens1993melting, wang1999properties}.

\subsection{Rogers-Young scheme}

The revised $\delta\gamma$ method and the PA scheme require $g(r)$ as the only input.  
We obtain $g(r)$ numerically by solving the Ornstein-Zernike equation \cite{Hansen_McDonald1986},
\begin{align}\label{equ:O-Z}
h(r) = c(r) + n \int \!\! \text{d}^3 r' c(r') h(|{\bf r}-{\bf r}'|)  
\end{align}
for a three-dimensional, homogeneous and isotropic fluid in conjunction with the approximate RY \cite{Rogers1984} closure relation invoking the HSY pair potential,  
\begin{align}\label{equ:Rogers-Young}
\frac{u(r)}{k_B T} + \ln g(r) = \ln\left[1 + \frac{\exp\{[h(r)-c(r)] f(r)\}-1}{f(r)} \right]\,.  
\end{align}
Here, $h(r) = g(r) - 1$ is the total correlation function, $c(r)$ is the direct correlation function,
and $n$ is the particle number density.
Eq.~\eqref{equ:Rogers-Young} includes the mixing function $f(r) = 1 - \exp\{-\alpha r\}$ with the non-negative  
inverse length parameter $\alpha$. This parameter is determined self-consistently by requiring equality of the isothermal osmotic compressibilities derived from compressibility equation,  
\begin{align}\label{equ:invcomp_fluct}
{\left( \frac{\partial P_c / (k_{\mathrm{B}}T) }{\partial n} \right)}_{T} = 1 - 4\pi n \int\limits_0^\infty dr\;\! r^2\;\!c(r)
\end{align}
and the numerically differentiated virial pressure equation,
\begin{eqnarray}\label{equ:invcomp_vir}
\frac{P_v}{k_{\mathrm{B}} T} = n &+& \frac{2 \pi}{3} n^2 \left\lbrace \sigma^3 g(\sigma^+)
- \frac{1}{k_B T} \int\limits_\sigma^\infty dr~r^3~g(r)~\frac{d u(r)}{d r}\right\rbrace\,,
\end{eqnarray}
where $P_c$ and $P_v$ is the isothermal osmotic pressure in the compressibility and virial equation, respectively. 

In using Eq. (\ref{equ:invcomp_vir}), we neglect any thermodynamic state dependence of the pair potential. 
As noted further up this section already, a consequence of integrating out the microionic and solvent degrees of freedom is that 
the resulting effective pair potential of HSY form is in general dependent on the particle concentration $n$ and the system temperature $T$ (see, e.g., \cite{HeinenPalbergLoewen2014}). This gives rise to additional terms on the right-hand-side of Eq. (\ref{equ:invcomp_vir}) invoking the partial derivative of $u(r)$ with respect to $n$ and $T$. The precise form of the effective pair potential depends on the specific electro-chemical surface properties of the colloidal spheres, and the specific properties of the suspending electrolyte solution. Since we are not dealing here with the microscopic theory of effective colloidal pair potentials but with the generic behavior of $D^r$, we disregard any specific state dependence of $u(r)$, and of the RY mixing parameter $\alpha$.  

Our numerical solution of the RY integral equation for broad ranges of volume fractions, screening and interaction parameters has been facilitated by using a spectral solver described in Ref.~\cite{Heinen2014}. 
The good accuracy of the RY approximation for HSY systems was demonstrated in various 
studies \cite{nagele1996dynamics,Gapinski2005,banchio2008short,gapinski2009structure,heinen2011pair,Heinen2011Erratum}
comprising comparisons with simulation and experimental data.

\subsection{Radial distribution function in the RY scheme}\label{subsec:RDF}

Owing the rather steep ${\cal O}(1/r^6)$ long-distance decay of the rotational mobility tensor $\boldsymbol{\mu }_{ii}^{rr}$ associated with $D^r$ (see Sec. \ref{sec:PA}), the rotational diffusion coefficient is quite sensitive to the shape of the RDF at small particle separations. This motivates the following discussion on the behavior of $g(r)$ in the $\left(\lambda,\tilde{T},\phi\right)$ fluid-phase parameter regime of the HSY model explored in this work. 

\begin{figure}
\begin{centering}
\includegraphics[width=.6\textwidth]%
{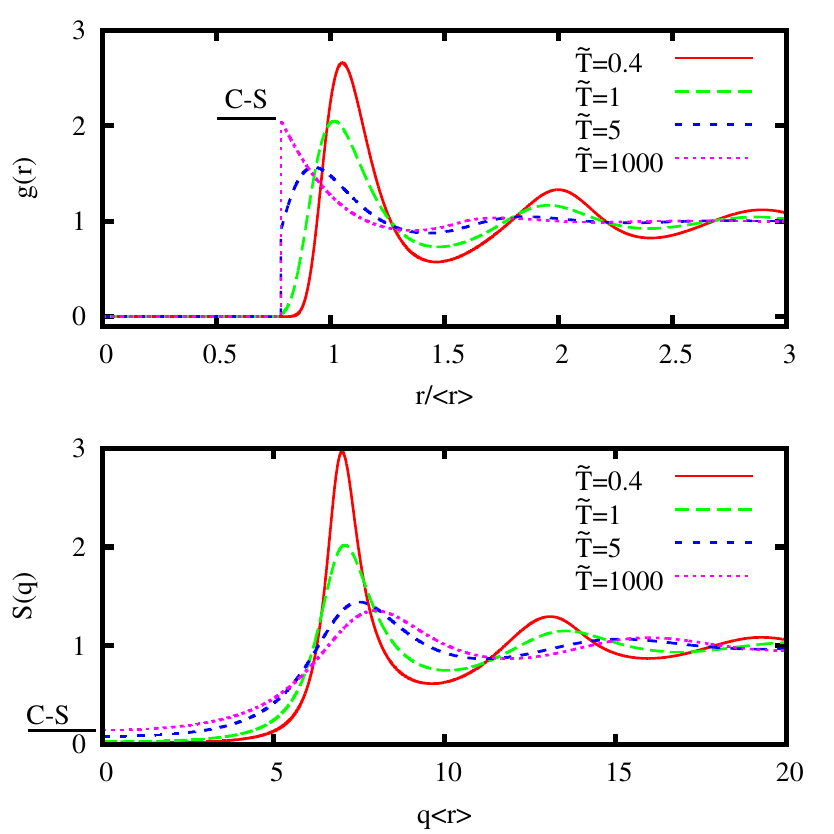}%
\vspace{4em}
\caption{(Color online) Upper panel: RDF, $g\left(r\right)$, predicted by the 
Rogers-Young scheme for a HSY fluid system with $\lambda=8$ and $\phi=0.25$. Various inverse Yukawa interaction  parameters 
$\tilde{T}=0.4,1,5,1000$ are considered as indicated. The horizontal dashed line segment marks the Carnahan-Starling (C-S)  contact value of hard spheres given by Eq. (\ref{radial at sigma CS}). Lower panel: Static 
structure factor, $S\left(  q\right)$, corresponding to the displayed radial
distribution functions in the upper panel. Dashed horizontal line segment: C-S compressibility factor of hard spheres according to Eq. (\ref{Sqzero CS}). Pair distance $r$ and wavenumber $q$ are scaled with the geometric mean particle distance $\langle r \rangle$.}%
\label{fig a few g(r) and S(q)}%
\end{centering}
\end{figure}

In Fig.~\ref{fig a few g(r) and S(q)}, the RY calculated RDFs (upper panel) and structure factors (lower panel) 
for $\phi=0.25$ 
and $\lambda=8$ are depicted for different inverse Yukawa interaction parameters $\tilde
{T}$ as indicated in the figure. The strength of the Yukawa potential at mean particle distance, 
in units of the thermal energy, decreases with increasing $\tilde{T}$. For the largest considered value  
$\tilde{T}=1000$, the HSY system at $\phi=0.25$ reduces essentially to a hard-sphere fluid, 
with the RDF maximum $g(r_m)$ located at contact distance $r_m=\sigma$. With decreasing
$\tilde{T}$, the increasingly strong Yukawa repulsion reduces the relative probability, $g(\sigma^+)$, of two-sphere contact, and it also lowers the compressibility factor $S(0)$. 
Moreover, the nearest neighbor shell of spheres around the radial distance $r_m$ where $g(r)$ has its maximum moves  outwards and sharpens with decreasing $\tilde{T}$. For the lowest considered value $\tilde{T}=0.4$ corresponding to a fluid state point near to the fluid-fcc phase boundary line, the hard core of the particles is masked and $r_m \approx \langle r \rangle$. 

\begin{figure}
\begin{centering}
\includegraphics[width=.6\textwidth]%
{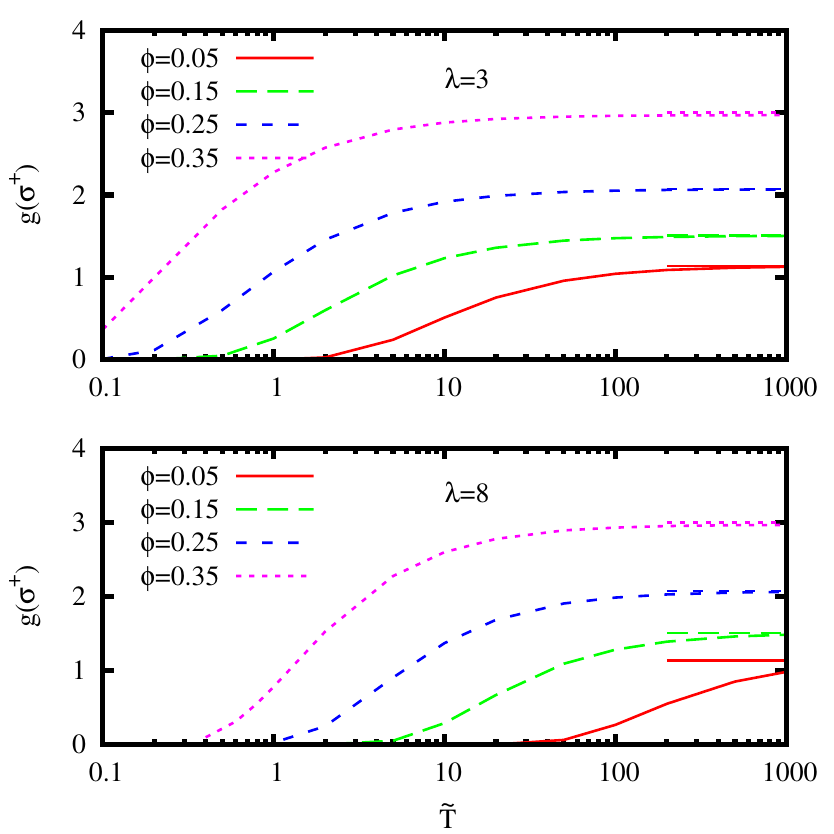}%
\vspace{4em}
\caption{(Color online) Contact value, $g\left(\sigma^+\right)$, of the HSY-RY $g(r)$ 
as a function of $\tilde{T}$, for volume fractions $\phi
=0.05,0.15,0.25,0.35$ as indicated. 
Upper panel: $\lambda=3$. Lower panel: $\lambda=8$.
Horizontal line segments for large $\tilde{T}$ mark the C-S hard-sphere contact values 
according to Eq. (\ref{radial at sigma CS}).}
\label{fig radial at contact}%
\end{centering}
\end{figure}

A measure of the importance of the hard-core part of $u(r)$ relative to the Yukawa part is given 
by the RDF contact value  $g\left(\sigma^+\right)$ plotted in Fig. 
\ref{fig radial at contact} for all considered fluid-phase points $(\lambda, 
\tilde{T},\phi)$.%
With increasing $\tilde{T}$, the relative strength of the Yukawa potential part ceases, and a plateau region of the RDF contact value is approached, characteristic for hard-sphere-like behavior. This is obviated by the horizontal line segments shown at the right ordinate of the figure which mark the Carnahan-Starling (C-S) RDF contact values of hard spheres with vanishing Yukawa tail repulsion ($\gamma = 0$), given by
\begin{equation}
	g_{HS}^{CS}\left(  \sigma^{+}\right)  =\frac{1-\frac{1}{2}\phi}{\left(
1-\phi\right)  ^{3}}\,. \label{radial at sigma CS}%
\end{equation}
We quote in addition the C-S equation,
\begin{equation}
S_{HS}^{CS}\left(q=0\right) = \frac{\left(1-\phi\right)^4}{\left(1+2\phi\right)^2+\phi^3\left(\phi-4\right)} \,,
\label{Sqzero CS}%
\end{equation}
for the compressibility factor of hard spheres.
The origin of  
the excellent accuracy of the semi-phenomenological C-S expressions for hard spheres is still a riddle, 
and a topic of ongoing research \cite{Robles2014}.
We emphasize that the employed RY scheme is a quite accurate but
nevertheless approximate integral equation. Its partial
thermodynamic self consistency does not imply, e.g.,  
perfect agreement of the RY contact value for hard spheres with the practically exact 
Carnahan-Starling result in Eq.~\eqref{radial at sigma CS}. In fact, the RY
scheme is lacking thermodynamic self-consistency with respect to any
thermodynamic property except for the pressure. An extended version of
the RY scheme (named ERY scheme) has been introduced by Carbajal-Tinoco
\cite{CarbajalTinoco2008}, which further improves the accuracy
of the original RY scheme by introducing a second mixing parameter. In
this more elaborate scheme, which however is not applicable to pure hard spheres without soft repulsion,  
the two mixing parameters are determined by enforcing thermodynamic self-consistency both regarding 
the pressure and the excess internal energy per particle. For
the sake of simplicity and numerical stability, we have refrained from using
the ERY scheme in the present work. 
 
On first sight, it is surprising that in accordance with Fig. \ref{fig radial at contact}, the likelihood of observing two closely spaced  particles is larger for $\lambda=3$ than for $\lambda=8$, for equal $\phi$ and $\tilde{T}$, even though the screening length of the Yukawa tail is significantly shorter in the latter case. This can be understood as follows: While the potential value $\beta u(x=1)=1/\tilde{T}$ at $r=\langle r \rangle$ is equal for both $\lambda$ values, the repulsive force, $-\beta d\;\!u/dx(x=1)= \left(1 +\lambda\right)/\tilde{T}$, is larger in the $\lambda=8$ case. Taken together with the substantially steeper rise of the Yukawa potential with decreasing $x$ for $\lambda=8$, this explains the lower probability of finding two closely spaced particles. For fixed $\lambda$, 
the small-$\tilde{T}$ region where the hard-core is masked (i.e. where $g(\sigma^+) \approx 0$) 
shrinks with increasing $\phi$, as it is expected. 

For the upcoming discussion of $D^r$, it is relevant to investigate how the principal RDF maximum 
$g(r_m)$, and its position $r_{m}$, depend on the pair potential parameters. 
We notice first from Fig.~\ref{fig a few g(r) and S(q)} that $r_m$ equals the smallest radial 
distance $r \geq \sigma$ where the derivative of $g(r)$ turns negative. 

\begin{figure}
\begin{centering}
\includegraphics[width=.6\textwidth]%
{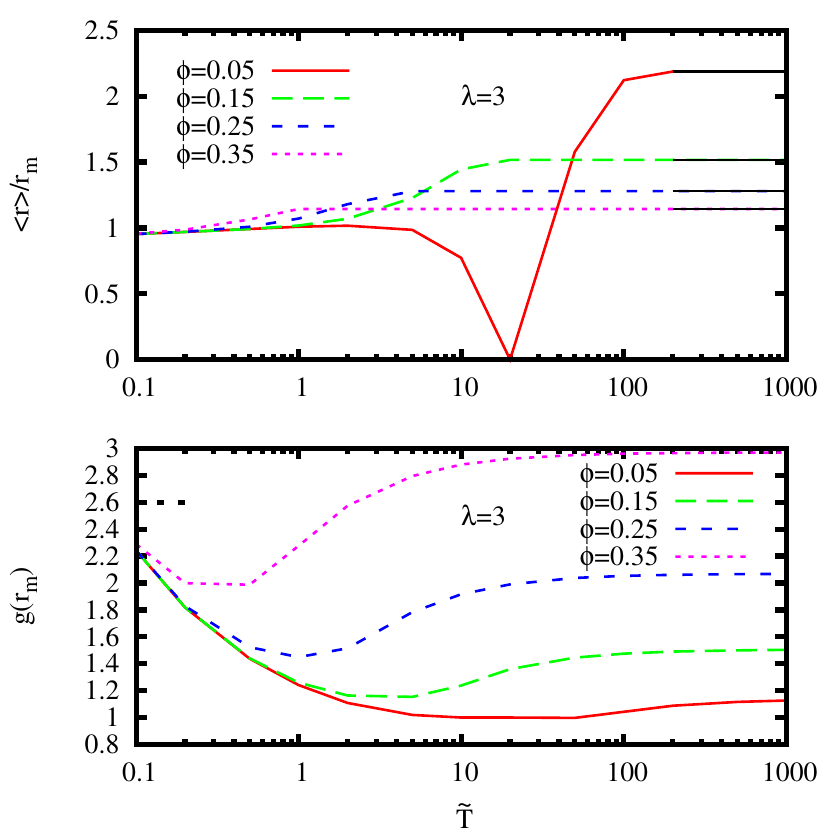}%
\vspace{4em}
\caption{(Color online) Upper panel: Inverse reduced position, $\langle r \rangle/r_{m}$, 
of the principal peak of $g(r)$ as a function of $\tilde{T}$, for $\lambda=3$ and values of $\phi$ as indicated. The horizontal line segments
at large $\tilde{T}$ indicate the inverse reduced contact distance, $\langle r \rangle/\sigma$, for the respective $\phi$ values. Lower panel: Principal peak value, $g\left(r_{m}\right)$, of the RDF for the same set of parameters. The dashed horizontal line segment at small $\tilde{T}$ indicates the one-component plasma isochoric freezing transition value 
attained in the zero-screening limit $\lambda \to 0$ 
(see \cite{gapinski2014freezing}).}%
\label{fig princ max lambda 3}%
\end{centering}
\end{figure}
\begin{figure}
\begin{centering}
\includegraphics[width=.6\textwidth]%
{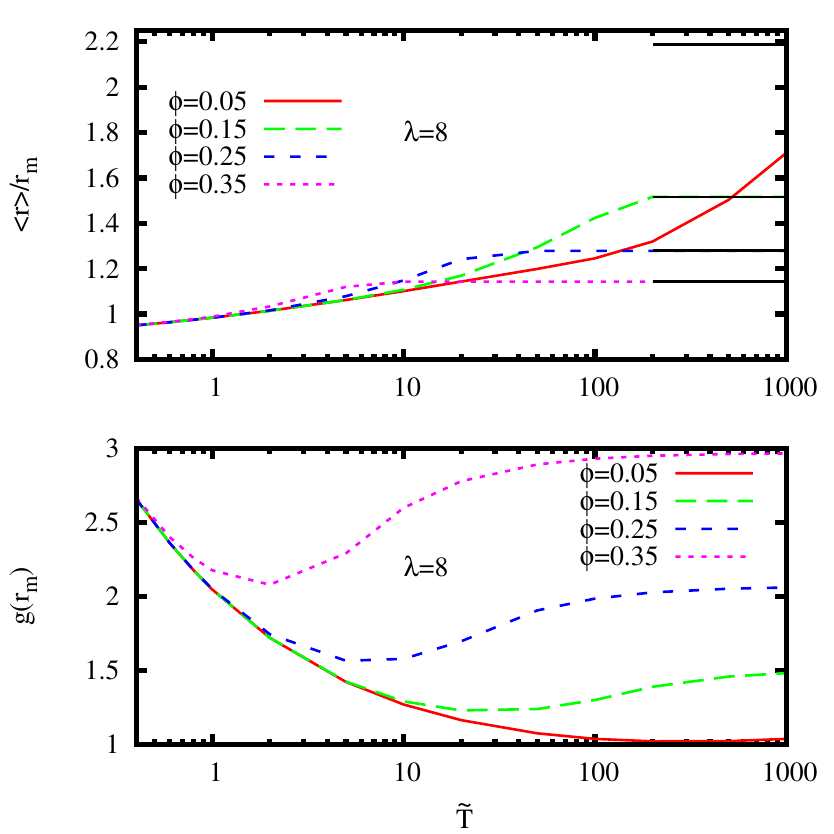}%
\vspace{4em}
\caption{(Color online) Upper and lower panels: Same as in the upper and lower panel of Fig. \ref{fig princ max lambda 3}, respectively, but for $\lambda=8$.}%
\label{fig princ max lambda 8}%
\end{centering}
\end{figure}
The dependence of $g(r_m)$ and $r_m$ on $\tilde{T}$ is shown 
Fig.~\ref{fig princ max lambda 3} for $\lambda=3,$ and in Fig.~\ref{fig princ max lambda 8} for $\lambda=8$. 
According to both figures, for fixed $\phi$ and therefore fixed mean particle distance $\langle r \rangle$, the inverse principal peak location, $\langle r \rangle/r_m$, increases with decreasing Yukawa potential strength, i.e. increasing $\tilde{T}$, towards the limiting inverse reduced contact distance, $\langle r \rangle/\sigma$, of neutral hard spheres. The limiting hard-sphere values for the considered $\phi$ are indicated by the horizontal solid line segments at the 
large-$\tilde{T}$ end of the upper panels in Figs.~\ref{fig princ max lambda 3} and \ref{fig princ max lambda 8}. Except for $\lambda=8$ and the lowest considered volume fraction $\phi=0.05$, the hard-sphere limiting values have been all reached for $\tilde{T}=1000$. It is for this $(\lambda,\phi)$ point where the minimum of $g(r_m)$ as a function of $\tilde{T}$ observed in all depicted curves in the lower panels of Figs. (\ref{fig princ max lambda 3}) and (\ref{fig princ max lambda 8}) has its largest $\tilde{T}$ value. The minimum in $g(r_m)$ originates from the competition of Yukawa repulsion and excluded volume interaction (see here again Fig.~\ref{fig a few g(r) and S(q)} for $g(r)$). This competition is enforced  with  increasing concentration, reflected by a more pronounced minimum moving inwards to smaller values of $\tilde{T}$. 

\begin{figure}
\begin{centering}
\includegraphics[width=.6\textwidth]%
{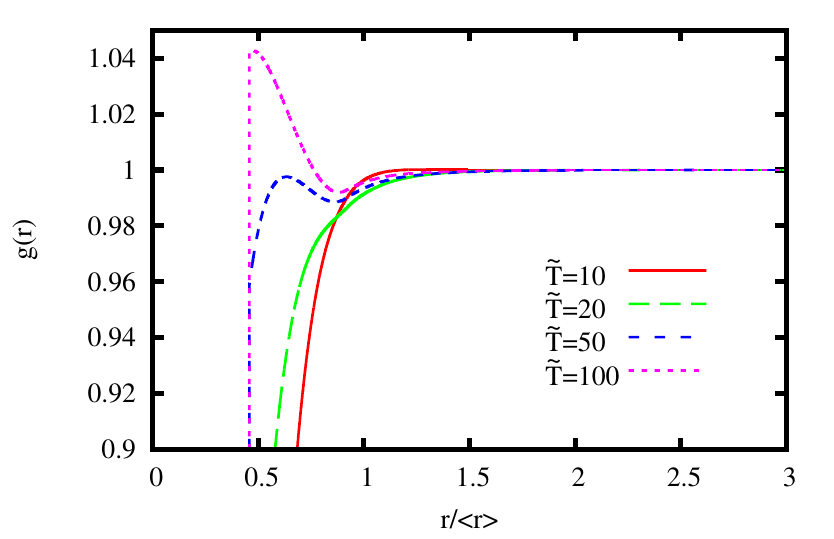}%
\vspace{4em}
\caption{(Color online) HSY-RY $g\left(r\right)$  
for $\phi=0.05$ and $\lambda=3$. Employed values of $\tilde{T}$ 
are $\tilde{T}=10,20,50,100$ as indicated.}
\label{fig a few radial for lambda 3 phi 0.05}%
\end{centering}
\end{figure}

We finally notice that the only exception from the monotonic behavior of
$r_{m}$ as a function of $\tilde{T}$ is the curve in the upper panel of Fig.~\ref{fig princ max lambda 3} for
$\lambda=3$ and $\phi=0.05$. To explain the peculiar shape of this curve, 
in Fig.~\ref{fig a few radial for lambda 3 phi 0.05} we plot associated RDFs,   
for values of $\tilde{T}$ as indicated. 
For large $\tilde{T}\geq50$, the principal maximum of $g(r)$ is located close to $r=\sigma$.
The maximum decreases and is shifted to larger distances $r_m$ with decreasing $\tilde{T}$. 
At $\tilde{T}=20$, no localized principal maximum is present any more, and $g(r)$ is 
monotonically increasing with increasing $r$ so that $r_m= \infty$. When $\tilde{T}$ is further decreased, 
the strengthened Yukawa repulsion causes the reappearance of a RDF maximum   
at a distance $r_m$ significantly larger than $\sigma$ which is decreasing towards $\langle r \rangle$.

\section{Revised Beenakker-Mazur method} \label{sec:Revised-BM}

We present here the details of our revised $\delta\gamma$ method in its application to rotational self-diffusion. 
The calculation of short-time rotational self-diffusion coefficients on basis of Eq.~(\ref{rot self diff}) is not straightforward. The difficulty lies both in the calculation
of the equilibrium probability distribution function, and of the $3N\times 3N$ rotational mobility
matrix $\boldsymbol{\mu}^{rr}$. The latter difficulty is made explicit when the $3 \times 3$ tensor elements of the 
mobility matrix are represented in the form of a hydrodynamic scattering series,
\begin{equation}
\boldsymbol{\mu}_{ij}^{rr}\left(  \mathbf{R}_{1}\ldots\mathbf{R}_{N}\right)
=P^{r}T_{ij}\left(  \mathbf{R}_{1}\ldots\mathbf{R}_{N}\right)  P^{r},
\label{mu rr}%
\end{equation}
with $T_{ij}$ given by
\begin{eqnarray}
	T_{ij}\left(  \mathbf{R}_{1}\ldots\mathbf{R}_{N}\right)  & = & \delta_{ij}M\left(
\mathbf{R}_{i}\right)  +\left(  1-\delta_{ij}\right)  M\left(  \mathbf{R}%
_{i}\right)  G\left(  \mathbf{R}_{i},\mathbf{R}_{j}\right)  M\left(
\mathbf{R}_{j}\right)  + \nonumber \\
&&\sum_{\substack{k=1,\\k\neq i,k\neq j}}^{N}M\left(
\mathbf{R}_{i}\right)  G\left(  \mathbf{R}_{i},\mathbf{R}_{k}\right)  M\left(
\mathbf{R}_{k}\right)  G\left(  \mathbf{R}_{k},\mathbf{R}_{j}\right)  M\left(
\mathbf{R}_{j}\right)  +\ldots \,. \label{def T}%
\end{eqnarray}
The usage of scattering series in the context of suspensions is well established  \cite{smoluchowski1912practical,felderhof1990hydrodynamics}. Not to interrupt unnecessarily our line of reasoning, we therefore 
refer to the appendix for the explicit expressions for the matrices $M$ and $G$, and the definition
of the projection operator $P^r$ in Eq.~(\ref{mu rr}). It is worth while to note here, however, 
that the matrix $G$ is related to the Green
function tensor,
\begin{equation}
\mathbf{G}_{0}\left(  \mathbf{r}\right)  =\frac{1}{8\pi\eta r}\left(
\mathbf{1+\hat{r}\hat{r}}\right) \,,
\end{equation}
of the Stokes equations for an unbounded infinite fluid where $\mathbf{\hat{r}}=\mathbf{r}/r$. 
This so-called Oseen tensor describes the flow generated by a point force at the origin. Moreover,
the matrix $M\left(  \mathbf{R}_{i}\right)$ describes the hydrodynamic response of a single
sphere with its center at position $\mathbf{R}_{i}$. Each term in
Eq.~(\ref{def T}) is called a scattering sequence. Two examples of a scattering sequence are 
\begin{equation}
M\left(  \mathbf{R}_{1}\right)  G\left(  \mathbf{R}_{1},\mathbf{R}_{2}\right)
M\left(  \mathbf{R}_{2}\right)  G\left(  \mathbf{R}_{2},\mathbf{R}_{1}\right)
M\left(  \mathbf{R}_{1}\right)  G\left(  \mathbf{R}_{1},\mathbf{R}_{2}\right)
M\left(  \mathbf{R}_{2}\right)  \label{scattering sequence example 1}%
\end{equation}
and
\begin{equation}
M\left(  \mathbf{R}_{1}\right)  G\left(  \mathbf{R}_{1},\mathbf{R}_{3}\right)
M\left(  \mathbf{R}_{3}\right)  G\left(  \mathbf{R}_{3},\mathbf{R}_{2}\right)
M\left(  \mathbf{R}_{2}\right) \,. \label{scattering sequence example 2}%
\end{equation}
Each scattering sequence is a succession of matrices $M$ which scatter the
flow, and matrices $G$ which freely propagate the flow between scattering events. 
This physical interpretation of a scattering sequence is useful in our
further discussion, and for the description of the physical idea underlying the 
Beenakker-Mazur (BM) method.

\subsection{Fluctuation expansion}

In one of their first papers on short-time transport
properties of suspensions \cite{selfdiffusion83}, Beenakker and Mazur introduced the so-called fluctuation expansion.
The physical idea behind this expansion is related to the resummation of a certain
class of scattering sequences. BM based their considerations on the
scattering series given by the Eq. (2.2) in their Ref. 
\cite{selfdiffusion83}. In the present analysis, we formulate the fluctuation expansion on basis 
of the scattering series given by
Eq. (\ref{def T}). Both approaches are equivalent but the presentation of the BM 
theory is more straightforward using our scattering series representation.

\begin{figure}
\begin{centering}
\fbox{\includegraphics[width=.6\textwidth]%
{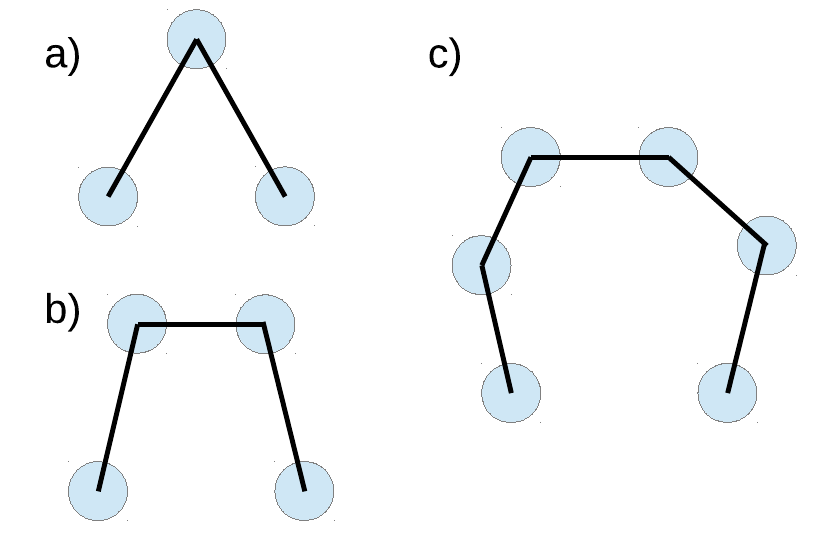}}%
\vspace{1em}
\caption{
Schematic representation of scattering sequences resummed in the fluctuation expansion.
a) Scattering sequence in Eq.~(\ref{two reflections fluct exp}).
b) Scattering sequence in Eq.~(\ref{three reflections fluct exp}).
c) Scattering sequence with five propagators.
}
\label{fig fluctuation expansion}%
\end{centering}
\end{figure}

The scattering sequence in Eq.~(\ref{scattering sequence example 2})
starts at particle $2$ at position ${\bf R}_2$ and ends at particle $1$ at position ${\bf R}_1$, 
with particle $3$ acting as the intermediate. 
The scattering sequence in Eq.~\eqref{def T}, with the first propagators $G$ 
starting from the particle $2$ and the second one ending at the
particle $1$, can have all the other $N-2$ particles as intermediates, summing thus up to the expression
\begin{equation}
\sum_{i=3}^{N}M\left(  \mathbf{R}_{1}\right)  G\left(  \mathbf{R}%
_{1},\mathbf{R}_{i}\right)  M\left(  \mathbf{R}_{i}\right)  G\left(
\mathbf{R}_{i},\mathbf{R}_{2}\right)  M\left(  \mathbf{R}_{2}\right)\,.
\end{equation}
If we treat this expression on the mean-field level by neglecting correlations
between the particle positions, it is approximated by the volume integral
\begin{equation}
n\int d^{3}R_{3}\;\!M\left(  \mathbf{R}_{1}\right)  G\left(  \mathbf{R}%
_{1},\mathbf{R}_{3}\right)  M\left(
\mathbf{R}_{3}\right)  G\left(  \mathbf{R}_{3},\mathbf{R}_{2}\right)  M\left(
\mathbf{R}_{2}\right)  \label{two reflections fluct exp}%
\end{equation}
with the number density $n$ playing in a homogeneous system the role of the single-particle distribution function.
The procedure of summing up similar scattering sequences involving three
propagators $G$ results in 
\begin{equation}
n^2 \int d^{3}R_{3}\int d^{3}\;\!R_{4}M\left(  \mathbf{R}_{1}\right)  G\left(
\mathbf{R}_{1},\mathbf{R}_{3}\right)   M\left(
\mathbf{R}_{3}\right)  G\left(  \mathbf{R}_{3},\mathbf{R}_{4}\right)
M\left(  \mathbf{R}_{4}\right)  G\left(  \mathbf{R}%
_{4},\mathbf{R}_{2}\right)  M\left(  \mathbf{R}_{2}\right)  .
\label{three reflections fluct exp}%
\end{equation}
Notice that in the scattering sequences described by Eqs.
(\ref{two reflections fluct exp}) and (\ref{three reflections fluct exp}),
every reflection in the sequence is directed towards a new particle. 
This is schematically 
shown in Fig.~\ref{fig fluctuation expansion}.
An example of a scattering sequence not satisfying this condition is given
by Eq.~(\ref{scattering sequence example 1}).

To proceed further, it is convenient to follow BM 
by introducing integral density kernels. Therefore, instead of $T_{ij}\left(
\mathbf{R}_{1}\ldots\mathbf{R}_{N}\right)$ in Eq. (\ref{def T}), we
consider its kernel density $\mathcal{T}$ defined by
\begin{equation}
\mathcal{T}\left(  \mathbf{R},\mathbf{R}^{\prime};\mathbf{R}_{1}%
\ldots\mathbf{R}_{N}\right)  \mathcal{=}\sum_{i,j=1}^{N}\delta\left(
\mathbf{R}-\mathbf{R}_{i} \right)  T_{ij}\left(  \mathbf{R}_{1}\ldots\mathbf{R}_{N}\right)
\delta\left(  \mathbf{R}^{\prime}-\mathbf{R}_{j}\right)\,,
\end{equation}
where $\delta({\bf R})$ is the three-dimensional delta function. 
The kernel $\mathcal{T}$ is represented now as follows,
\begin{equation}
\mathcal{T=M+M}\tilde{G}\mathcal{M}+\mathcal{M}\tilde{G}\mathcal{M}\tilde
{G}\mathcal{M+\ldots} \label{scattering series T kernel}%
\end{equation}
where the densities $\mathcal{M}\left(  \mathbf{R},\mathbf{R}^{\prime};\mathbf{R}_{1}%
\ldots\mathbf{R}_{N}\right)  $ and $\tilde{G}\left(  \mathbf{R},\mathbf{R}%
^{\prime}\right)  $ are defined by
\begin{equation}
\mathcal{M}\left(  \mathbf{R},\mathbf{R}^{\prime};\mathbf{R}_{1}%
\ldots\mathbf{R}_{N}\right)  =\delta\left(  \mathbf{R}-\mathbf{R}^{\prime
}\right)  \sum_{i=1}^{N}M\left(  \mathbf{R}_{i}\right)  \delta\left(
\mathbf{R-R}_{i}\right)\,, \label{M kernel}%
\end{equation}
and
\begin{equation}
\tilde{G}\left(  \mathbf{R},\mathbf{R}^{\prime}\right)  =\left\{
\begin{array}
[c]{cc}%
\displaystyle
0 & \text{for }\mathbf{R}=\mathbf{R}^{\prime}\\%
\displaystyle
G\left(  \mathbf{R},\mathbf{R}^{\prime}\right)  & \text{for }\mathbf{R}%
\not =\mathbf{R}^{\prime}\text{ }%
\end{array}
\right. \,,
 \label{eq:G wave kernel}%
\end{equation}
respectively. 
In Eq.~(\ref{scattering series T kernel}), products of kernels such as 
$\mathcal{M}\tilde{G}$ appear which should be interpreted as:
\begin{equation}
\left[  \mathcal{M}\tilde{G}\right]  \left(  \mathbf{R},\mathbf{R}^{\prime
};\mathbf{R}_{1}\ldots\mathbf{R}_{N}\right)  \mathcal{\equiv}\int
d\mathbf{R}^{\prime\prime}\mathcal{M}\left(  \mathbf{R},\mathbf{R}%
^{\prime\prime};\mathbf{R}_{1}\ldots\mathbf{R}_{N}\right)  \tilde{G}\left(
\mathbf{R}^{\prime\prime},\mathbf{R}^{\prime}\right)\,.
\label{product of two kernels}%
\end{equation}
Notice that Eq. (\ref{scattering series T kernel}) is a 
short-hand notation since the kernel variables, the integral signs and the position vectors of the 
particles are omitted. The full notation is used, in contrast, in Eq.
(\ref{product of two kernels}). It is worth noting that the only
difference between $G\left(  \mathbf{R},\mathbf{R}^{\prime}\right)  $ and
$\tilde{G}\left(  \mathbf{R},\mathbf{R}^{\prime}\right)  $ in Eq.~\eqref{eq:G wave kernel}
is at $\mathbf{R}=\mathbf{R}^{\prime}$.
In introducing $\tilde{G}$ instead of $G$, BM have avoided the 
summation over the same particles since terms such as $M\left(  \mathbf{R}_{i}\right)
\tilde{G}\left(  \mathbf{R}_{i},\mathbf{R}_{i}\right)  M\left(  \mathbf{R}%
_{i}\right)$ are zero. The vanishing of $\tilde{G}\left(  \mathbf{R}_{i},\mathbf{R}%
_{j}\right)$ for $i =j$ excludes reflections to the same particle which are absent in the
initial scattering series in Eq. (\ref{def T}).

The sequence in Eq.~(\ref{two reflections fluct exp}) is expressed
in terms of the introduced kernels by
\begin{equation}
M\left(  \mathbf{R}_{1}\right)  \left[  \tilde{G}\left\langle \mathcal{M}%
\right\rangle \tilde{G}\right]  \left(  \mathbf{R}_{1},\mathbf{R}_{2}\right)
M\left(  \mathbf{R}_{2}\right)\,,
\end{equation}
and the sequence in Eq.~(\ref{three reflections fluct exp}) by
\begin{equation}
M\left(  \mathbf{R}_{1}\right)  \left[  \tilde{G}\left\langle \mathcal{M}%
\right\rangle \tilde{G}\left\langle \mathcal{M}\right\rangle \tilde{G}\right]
\left(  \mathbf{R}_{1},\mathbf{R}_{2}\right)  M\left(  \mathbf{R}_{2}\right)\,.
\end{equation}
As throughout this paper, $\langle \dots \rangle$ denotes the average with respect to the 
equilibrium configurational probability density function 
of $N$ particles.
Resummation of all scattering sequences for which each propagator connects to a different particle such as in 
the schematic scattering sequences in Fig.~\ref{fig fluctuation expansion},
yields
\begin{equation}
M\left(  \mathbf{R}_{1}\right)  G_{\left\langle \mathcal{M}\right\rangle
}\left(  \mathbf{R}_{1},\mathbf{R}_{2}\right)  M\left(  \mathbf{R}_{2}\right)\,.
\end{equation}
Here $G_{\left\langle \mathcal{M}\right\rangle }$ is an "effective propagator", 
defined by 
\begin{equation}
G_{\left\langle \mathcal{M}\right\rangle }=\tilde{G}\left(  1-\left\langle
\mathcal{M}\right\rangle \tilde{G}\right)  ^{-1}\,,
\end{equation}
which is the sum of a geometric series. 
The resummation of the above class of scattering sequences in the scattering series in Eq.~(\ref{scattering series T kernel})
leads to the so-called fluctuation expansion.
The resummation can be performed by rewriting Eq.~(\ref{scattering series T kernel}) as follows
\begin{equation}
\mathcal{T}=\mathcal{M+M}\tilde{G}\left[  1-\mathcal{M}\tilde{G}\right]
^{-1}\mathcal{M} \,.
\label{eq: T by  geometric series}%
\end{equation}
By adding and subtracting $\left\langle \mathcal{M}\right\rangle$, 
and using the operator identity 
\begin{equation}
\left(  1-A-B\right)
^{-1}=\left(  1-A\right)  ^{-1}\left[  1-B\left(  1-A\right)  ^{-1}\right]\,, 
\end{equation}
for two operators $A$ and $B$, 
we obtain the fluctuation expansion of the propagator $\tilde{G}\left[  1-\mathcal{M}\tilde{G}\right]  ^{-1}$ as
\begin{align}
\tilde{G}\left[  1-\mathcal{M}\tilde{G}\right]  ^{-1} &  =\tilde{G}\left[
1-\left(  \mathcal{M-}\left\langle \mathcal{M}\right\rangle \right)  \tilde
{G}-\left\langle \mathcal{M}\right\rangle \tilde{G}\right]  ^{-1}\\
&  =\tilde{G}_{\left\langle \mathcal{M}\right\rangle }\left[  1-\left(
\mathcal{M-}\left\langle \mathcal{M}\right\rangle \right)  \tilde
{G}_{\left\langle \mathcal{M}\right\rangle }\right]  ^{-1}\,.
\end{align}

When this expression is inserted in Eq.~\eqref{eq: T by  geometric series},
the fluctuation expansion result 
\begin{equation}
\mathcal{T}=\mathcal{M+M}\tilde{G}_{\left\langle \mathcal{M}\right\rangle
}\left[  1-\left(  \mathcal{M-}\left\langle \mathcal{M}\right\rangle \right)
\tilde{G}_{\left\langle \mathcal{M}\right\rangle }\right]  ^{-1}%
\mathcal{M}\,,
\label{fluctuation expansion}%
\end{equation}
for the kernel $\mathcal{T}$ is obtained.

\subsection{Renormalized fluctuation expansion}

The truncation of the fluctuation expansion for the translational mobility matrix has led to an approximate method of calculating the 
short-time translational self-diffusion coefficients \cite{selfdiffusion83}. However, the results by this  
second-order fluctuation expansion were found by BM to be unsatisfactory.  
Therefore, in subsequent work, BM performed another resummation which has  
led to the renormalized fluctuation expansion \cite{resummation83}.

\begin{figure}
\begin{centering}
\fbox{\includegraphics[width=.6\textwidth]%
{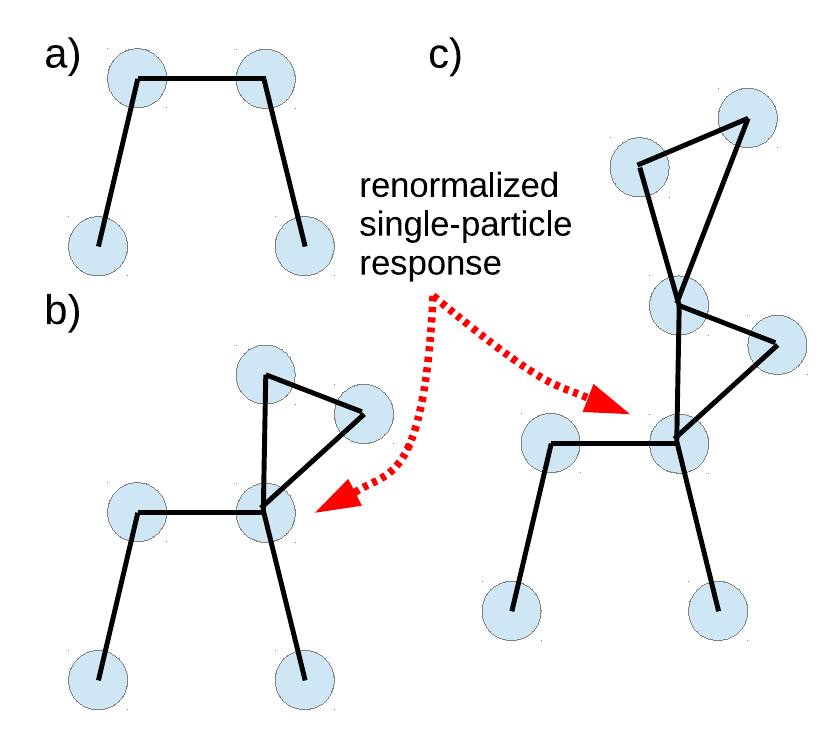}}%
\vspace{1em}
\caption{
Schematic representation of scattering sequences resummed in a renormalized fluctuation expansion 
invoking the renormalized propagator in Eq. (\ref{gmr}). 
a) Scattering sequence without renormalization of the single-particle response (present also in the fluctuation expansion). 
b) and c) Scattering sequence with the single-particle operator $M$ being renormalized.
}
\label{fig renormalized fluctuation expansion}%
\end{centering}
\end{figure}

The scattering sequences illustrated in Fig.~\ref{fig fluctuation expansion} for which each propagator links two different particles are resummed
in the fluctuation expansion. In the renormalized
fluctuation expansion, BM resummed similar scattering sequences
using now a renormalized single-particle operator $M_R$. Namely, instead of the bare $M$
operators, there appear now 'rings' built of scattering sequences with
renormalized operators. This is illustrated schematically in Fig.~\ref{fig renormalized fluctuation expansion}. 
BM refer to these structures as 'ring self-correlations',
which are scattering structures of the following form
\begin{equation}
\mathcal{M}_{R}=\mathcal{M}\left(  1-G_{\left\langle \mathcal{M}%
_{R}\right\rangle }^{s}\mathcal{M}\right)  ^{-1}, \label{mr}%
\end{equation}
where the integral operator $G_{\left\langle \mathcal{M}_{R}\right\rangle }^{s}$
is defined by
\begin{equation}
G_{\left\langle \mathcal{M}_{R}\right\rangle }^{s}\left(  \mathbf{R,R}%
^{\prime}\right)  =\left\{
\begin{array}
[c]{cc}%
\displaystyle
G_{\left\langle \mathcal{M}_{R}\right\rangle }\left(  \mathbf{R}%
,\mathbf{R}^{\prime}\right)  & \text{for }\mathbf{R}=\mathbf{R}^{\prime}\\%
\displaystyle
0 & \text{for }\mathbf{R}\not =\mathbf{R}^{\prime}\text{ }%
\end{array}
\right.  , \label{gmrs}%
\end{equation}
while the renormalized propagator $G_{\left\langle \mathcal{M}_{R}\right\rangle
}$ is of the form
\begin{equation}
G_{\left\langle \mathcal{M}_{R}\right\rangle }=\tilde{G}\left(  1-\left\langle
\mathcal{M}_{R}\right\rangle \tilde{G}\right)  ^{-1}\,. \label{gmr}%
\end{equation}

The derivation of the expansion in renormalized density fluctuations is omitted here 
for it involves only simple algebraic manipulations which are given in the works of 
BM \cite{resummation83,Beenakker1984effective,Beenakker1984Diffusion}. The most
important result of this derivation is the expansion of the operator $\mathcal{T}$  in
renormalized fluctuations, $\mathcal{M}_{R}-\left\langle \mathcal{M}%
_{R}\right\rangle $, according to
\begin{equation}
\mathcal{T}=\mathcal{M+M}G_{\left\langle \mathcal{M}_{R}\right\rangle }\left[
1-\left(  \mathcal{M}_{R}-\left\langle \mathcal{M}_{R}\right\rangle \right)
\tilde{G}_{\left\langle \mathcal{M}_{R}\right\rangle }\right]  ^{-1}%
\mathcal{M}_{R}\,. \label{rozw T we flukt gestosci}%
\end{equation}
The operator $\tilde{G}_{\left\langle \mathcal{M}_{R}\right\rangle }$ in 
the above expansion is defined by
\begin{equation}
\tilde{G}_{\left\langle \mathcal{M}_{R}\right\rangle }\left(  \mathbf{R}%
_{1},\mathbf{R}_{2}\right)  =G_{\left\langle \mathcal{M}_{R}\right\rangle
}\left(  \mathbf{R}_{1},\mathbf{R}_{2}\right)  -G_{\left\langle \mathcal{M}%
_{R}\right\rangle }^{s}\left(  \mathbf{R}_{1},\mathbf{R}_{2}\right)  .
\label{gmr falka}%
\end{equation}
It is also worth quoting the intermediate result
\begin{align}
\tilde{G}\left(  1-\mathcal{M}\tilde{G}\right)  ^{-1}%
\mathcal{M}  &  = G_{\left\langle \mathcal{M}%
_{R}\right\rangle }\left[  1-\left(  \mathcal{M}_{R}-\left\langle
\mathcal{M}_{R}\right\rangle \right)  \tilde{G}_{\left\langle \mathcal{M}%
_{R}\right\rangle }\right]  ^{-1}\left(  \mathcal{M}_{R}-\left\langle
\mathcal{M}_{R}\right\rangle \right)  \nonumber\\
&  + G_{\left\langle \mathcal{M}_{R}\right\rangle }\left[
1-\left(  \mathcal{M}_{R}-\left\langle \mathcal{M}_{R}\right\rangle \right)
\tilde{G}_{\left\langle \mathcal{M}_{R}\right\rangle }\right]  ^{-1}%
\left\langle \mathcal{M}_{R}\right\rangle\,,
\label{Geff M ren fluct exp}%
\end{align}
used in the renormalized fluctuation expansion of the translational and rotational self-diffusion coefficients discussed in the following.

\subsection{Renormalized fluctuation expansion for translational self-diffusion}

Before considering the renormalized fluctuation expansion of the rotational self-diffusion coefficient, 
we consider first translational self-diffusion which was investigated earlier by Beenakker and Mazur.
The short-time translational self-diffusion coefficient $D^{t}$ is expressed by the following formula
\begin{equation}
D^{t}=\frac{k_{B}T}{3}\lim_{\infty }\left[\text{Tr}\left\langle \frac{1}{N}%
\sum_{i=1}^{N}\boldsymbol{\mu }_{ii}^{tt}\left( \mathbf{R}_{1}\ldots \mathbf{%
R}_{N}\right) \right\rangle \right]\,,
\label{eq: tran self diff}
\end{equation}
in analogy with Eq.~\eqref{rot self diff} for the rotational self-diffusion coefficient. 
The above formula has been re-expressed by BM in terms of kernels according to
\begin{equation}
D^{t}=\frac{k_{B}T}{3}\text{Tr}\left(P^{t}dP^{t}\right), \label{trans self dif by d}%
\end{equation}
where the matrix $d$ is defined as follows:
\begin{equation}
d=n M\left(  \mathbf{R}\right)  +
M\left(  \mathbf{R}\right)  \lim_{\infty}\left\langle
\tilde{G}\left(  1-\mathcal{M}\tilde{G}\right)  ^{-1}\mathcal{M}\right\rangle
\left(  \mathbf{R,R}\right) \,. \label{d definition}%
\end{equation}
The operator $P^{t}$ whose explicit form is given in the appendix projects on the 
translational components of the matrix $d$. This should be compared with the Eq. (3.16) in \cite{selfdiffusion83}. 
The operator expression in Eq.~\eqref{Geff M ren fluct exp} 
inserted into the above expression results in the renormalized  
fluctuation expansion of $d$. 
Up to second order in the renormalized fluctuations, $\mathcal{M}_{R}-\left\langle \mathcal{M}_{R}\right\rangle$, 
this expansion reads
\begin{equation}
d=d^{\left(  0\right)  }+d^{\left(  1\right)  }+d^{\left(  2\right)  }%
+\ldots,\label{d lowest orders}%
\end{equation}
where
\begin{eqnarray}
d^{\left(  0\right)  } &=&nM+M\left[  G_{\left\langle \mathcal{M}_{R}%
\right\rangle }\left\langle \mathcal{M}_{R}\right\rangle \right]  \left(
\mathbf{R,R}\right)\,, \\
d^{\left(  1\right)  }&=& 0\,, \\
d^{\left(  2\right)  }&=&d_{2}^{\left(  2\right)  }+d_{3}^{\left(  2\right)  } \,,
\end{eqnarray}
and
\begin{eqnarray}
d_{2}^{\left(  2\right)  }&=&M\left[  G_{\left\langle \mathcal{M}%
_{R}\right\rangle }\left\langle \left(  \mathcal{M}_{R}-\left\langle
\mathcal{M}_{R}\right\rangle \right)  \tilde{G}_{\left\langle \mathcal{M}%
_{R}\right\rangle }\left(  \mathcal{M}_{R}-\left\langle \mathcal{M}%
_{R}\right\rangle \right)  \right\rangle \right]  \left(  \mathbf{R,R}\right) \,, \\
d_{3}^{\left(  2\right)  }&=&M\left[  G_{\left\langle \mathcal{M}%
_{R}\right\rangle }\left\langle \left(  \mathcal{M}_{R}-\left\langle
\mathcal{M}_{R}\right\rangle \right)  \tilde{G}_{\left\langle \mathcal{M}%
_{R}\right\rangle }\left(  \mathcal{M}_{R}-\left\langle \mathcal{M}%
_{R}\right\rangle \right)  \right\rangle \tilde{G}_{\left\langle
\mathcal{M}_{R}\right\rangle }\left\langle \mathcal{M}_{R}\right\rangle
\right]  \left(  \mathbf{R,R}\right)\,.\label{d second order 3}%
\end{eqnarray}
The truncation approximation,
\begin{equation}
	d \approx d^{\left(  0\right)  }+d^{\left(  1\right)  }+d^{\left(  2\right)  } \,,   \label{second order app}
\end{equation}
constitutes along with Eq.~(\ref{trans self dif by d}) and Eqs.~(\ref{d lowest orders}-\ref{d second order 3}) 
the second-order renormalized fluctuation expansion approximation 
for the short-time translational self-diffusion coefficient.

\subsection{Renormalized fluctuation expansion for rotational self-diffusion}

The extension of the renormalized fluctuation expansion method ($\delta\gamma$ scheme) to 
short-time rotational
self-diffusion was made first by Treloar and Masters
\cite{treloar1989short}. This extension is straightforward in our formalism, since 
Eq.~(\ref{rot self diff}) for $D^{r}$ is similar in structure as Eq. (\ref{trans self dif by d}) for $D^t$. 
The only difference appears in the invoked projectors, i.e. instead of the projector $P^{t}$ 
on the translational components the projector $P^{r}$ on the rotational components is used 
in Eq.~(\ref{rot self diff}). Since the  
renormalized fluctuation expansion has been introduced in the previous section on the general 
level of the $d$ matrix, the extension from translational to rotational self-diffusion is straightforward.%

\subsection{High-accuracy second-order renormalized fluctuation expansion}

Next, we point out the differences between the original second-order BM method of calculating $D^r$ 
by Treloar and Masters, and our revised second-order $\delta\gamma$ scheme. 
Approximations in the $\delta \gamma$ scheme are made in particular in two calculation steps. 
The first one is the truncation of the series in renormalized fluctuations 
to second order described in Eq.~\eqref{second order app}. Secondly, also the    
matrices $d^{\left(  0\right)  }, d^{\left(  2\right)  }$ are approximated 
since they relate to infinite dimensional hydrodynamic matrices such as $M$, $G$,
$G_{\left\langle \mathcal{M}_{R}\right\rangle }$,
or $G_{\left\langle \mathcal{M}_{R}\right\rangle }^{s}$ which for a numerical evaluation must be truncated (see the appendix).
In the works by BM on the hydrodynamic function and the high-frequency effective viscosity,
and in the work by Treloar and Masters on rotational diffusion,
due to technical difficulties severe truncations of these matrices have been made as discussed in detail in Ref.  \cite{makuch2012transport}.
In our revised method, we also truncate the infinite dimensional hydrodynamic matrices, 
but different from earlier works an extrapolation to infinite dimension has been included.  
The details of this extrapolation procedure are along the lines described in Ref. \cite{makuch2012transport} and are thus not repeated here.

\section{Simulation method}  \label{sec:sumulationmethod}

We have calculated $D^{r}$ to high precision for no-slip spheres using a hydrodynamic multipole method corrected for
lubrication~\cite{trojczastkowasamodyfuzja,CFHWB,CFSch:88,cichocki2000friction}, and encoded in the
HYDROMULTIPOLE program package \cite{trojczastkowasamodyfuzja}. The values for $D^{r}$ have been determined from 
equilibrium configuration averages using typically $N = 256$ spheres  
interacting with the HSY potential, and placed in a periodically
replicated cubic simulation box. At least 150 independent configurations for
each parameter set $\left(\lambda,\tilde{T},\phi\right)$ were
used. This has resulted in a statistical relative error of less than 0.001. As
reported in Ref. \cite{abade2011rotational}, the calculated values for
$D^{r}(N)$ using the periodic simulation box with $N$ particles are
not critically dependent on the system size. Therefore, no system size correction extrapolating to an 
infinitely large system is required as for short-time collective diffusion properties. 

\section{Pairwise additivity approximation} \label{sec:PA}

The rotational hydrodynamic mobility tensor $\boldsymbol{\mu}^{rr}_{ii}(\mathbf{R}_{1} \ldots\mathbf{R}_{N})$ of $N$ spherical particles in an infinite, quiescent fluid linearly relates the hydrodynamic torque $\mathbf{T}_i$ acting on a particle $i$ to its rotational velocity $\boldsymbol{\Omega}_i$. By disregarding three-body and higher order hydrodynamic cluster contributions, on can approximate the exact $N$-particle rotational hydrodynamic mobility tensor by a sum of two-particle contributions, 
\begin{eqnarray}
{\boldsymbol{\mu}^{rr}_{ii}(\mathbf{R}_{1} \ldots\mathbf{R}_{N})} \approx \mu_0^r \left[ \mathds{1} +
\sum_{n=1;n\neq i}^N \boldsymbol{\omega}_{11}^{rr}(\mathbf{R}_i-\mathbf{R}_n) \right]\,, \label{eq:mob_pa}
\end{eqnarray}
where $\mu_0^r=D_0^r/k_BT$ is the single-sphere rotational mobility coefficient and $\mathds{1}$ the three-dimensional unit tensor. The two-sphere tensor $\boldsymbol{\omega}_{11}^{rr}(\mathbf{R}_i-\mathbf{R}_n)$ describes the hydrodynamic self-interaction of sphere $i$ by means of flow reflections at a second sphere labeled by $n$, in the absence of the  $N-2$ other particles. This constitutes the pairwise additivity (PA) approximation where it is assumed that the HIs between two spheres are not disturbed by other ones. In principle, this assumption is justified for semi-dilute systems only.   

On exploiting the axial symmetry of the two-sphere problem, the two-sphere tensor can be split into longitudinal and transversal components, 
\begin{equation}
\boldsymbol{\omega}_{11}^{rr}(\mathbf{R}_i-\mathbf{R}_n) =
\alpha_{11}^{rr}(R_{in})\hat{\mathbf{R}}_{in}\hat{\mathbf{R}}_{in} +
\beta_{11}^{rr}(R_{in})\left[ \mathds{1} - \hat{\mathbf{R}}_{in}\hat{\mathbf{R}}_{in} \right]\,,
\label{eq:2bodymob_split}
\end{equation}
with $\hat{\mathbf{R}}_{in} = (\mathbf{R}_i-\mathbf{R}_n) / R_{in}$ and $R_{in} = |\mathbf{R}_i-\mathbf{R}_n|$.

In term of the scalar longitudinal and transversal functions $\alpha_{11}^{rr}(R)$ and $\beta_{11}^{rr}(R)$, the 
normalized short-time rotational self-diffusion coefficient is expressed in PA approximation by \cite{PhysRevE.52.2707, Watzlawek1997,zhang2002tracer}
\begin{equation}\label{Dr_PA_approx}
\frac{D^r}{D_0^r} =  1 + 8\phi \int\limits_1^{\infty} dx~x^2 g(x) \left[ \alpha_{11}^{rr}(x) + 2\;\!\beta_{11}^{rr}(x) \right ] \,,
\end{equation}
where $x=r/\sigma$ is the two-sphere distance in units of the sphere diameter $\sigma=2a$. 
The functions $\alpha_{11}^{rr}(r)$ and $\beta_{11}^{rr}(r)$ can be calculated recursively in the form of a power series in the reduced inverse pair distance $a/r$. For the no-slip hydrodynamic surface boundary condition employed in this work, the leading-order (far-field) contributions are  
\begin{eqnarray}\label{eq:alpha_beta_farfield}
\alpha_{11}^{rr}(r) &=&
-3 {\left(\frac{a}{r}\right)}^8 + \mathcal{O}\left( {\left(\frac{a}{r} \right)}^{10} \right),\label{eq:alpha_farfield}\\
\beta_{11}^{rr}(r) &=&
-\frac{15}{4} {\left(\frac{a}{r}\right)}^6 - \frac{39}{4} {\left(\frac{a}{r}\right)}^8 +
\mathcal{O}\left( {\left(\frac{a}{r} \right)}^{10} \right)\,,
\label{eq:beta_farfield}
\end{eqnarray}
with higher-order terms in the expansion given in \cite{jones1988mobility, kim1991microhydrodynamics}.
At near-contact distance $r \approx 2a$ where lubrication comes into play, the expansions in Eqs.~\eqref{eq:alpha_farfield} and \eqref{eq:beta_farfield} converge only slowly. 
In our numerical implementation of the PA method, we therefore use  
high-order series expansion results obtained by Jeffrey and Onishi \cite{jeffrey1984calculation}. 
On using the zero concentration hard-sphere RDF, $g(x) = \Theta(x -1)$, in Eq.~\eqref{Dr_PA_approx}, 
where $\Theta(x)$ is the unit step function, we have numerically checked that our
code precisely reproduces the first-order virial coefficient value $0.631$ in Eq.~(\ref{eq:virial-HS-second}). 
This demonstrates the high accuracy of the employed tabulated values for $\alpha_{11}^{rr}(r)$ and $\beta_{11}^{rr}(r)$ also at near-contact distances.

\section{Results and discussion} \label{sec:results}

\begin{figure}
\begin{centering}
\includegraphics[width=.6\textwidth]%
{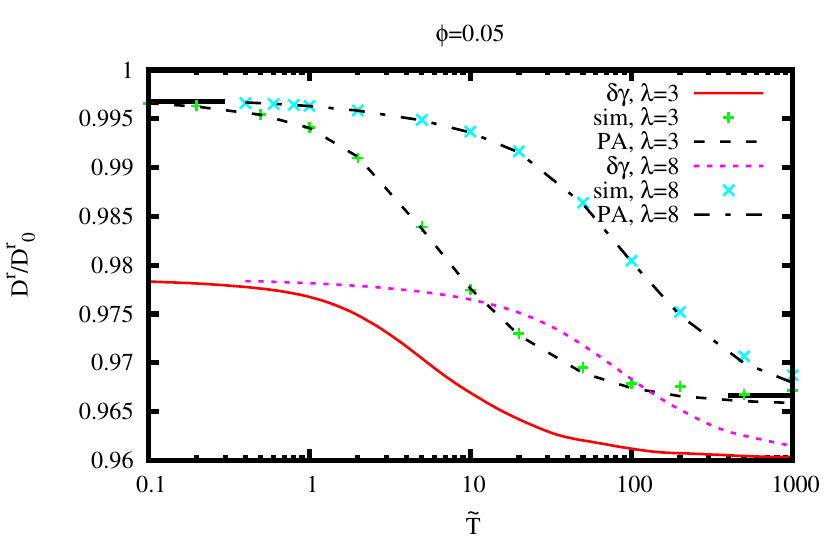}
\vspace{1em}
\caption{Normalized rotational self-diffusion coefficient, $D^{r}/D_{0}^{r}$, as a function of $\tilde{T}$, for $\lambda=3$ and $8$, and $\phi=0.05$. Simulation data (plus symbols and crosses) are compared with revised $\delta\gamma$ method predictions (solid and short-dashed curves) and PA method predictions (dashed-dotted and long-dashed curves). The theory and simulation values of $D^r$ for $\lambda=8$ are in general larger than the respective ones for $\lambda=3$. Horizontal solid segment at large $\tilde{T}$: Hard-sphere value according to Eq.~(\ref{eq:virial-HS-second}). 
Horizontal solid line segment at small $\tilde{T}$: Scaling prediction in Eq.~(\ref{eq:CSdeionized}) for low-salinity charge-stabilized systems.}
\label{fig dself rot 005}%
\end{centering}
\end{figure}

\begin{figure}
\begin{centering}
\includegraphics[width=.6\textwidth]%
{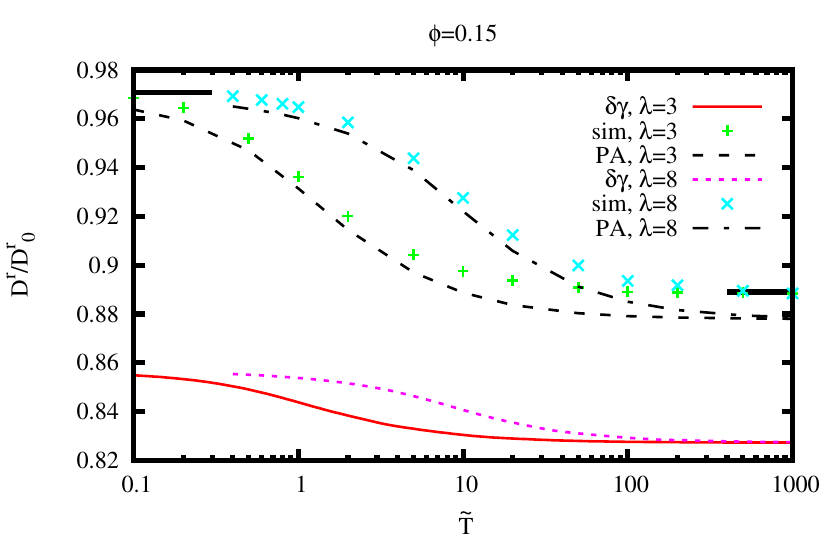}%
\vspace{1em}
\caption{Same as in Fig. \ref{fig dself rot 005} but for $\phi=0.15$.}%
\label{fig dself rot 015}%
\end{centering}
\end{figure}

\begin{figure}
\begin{centering}
\includegraphics[width=.6\textwidth]%
{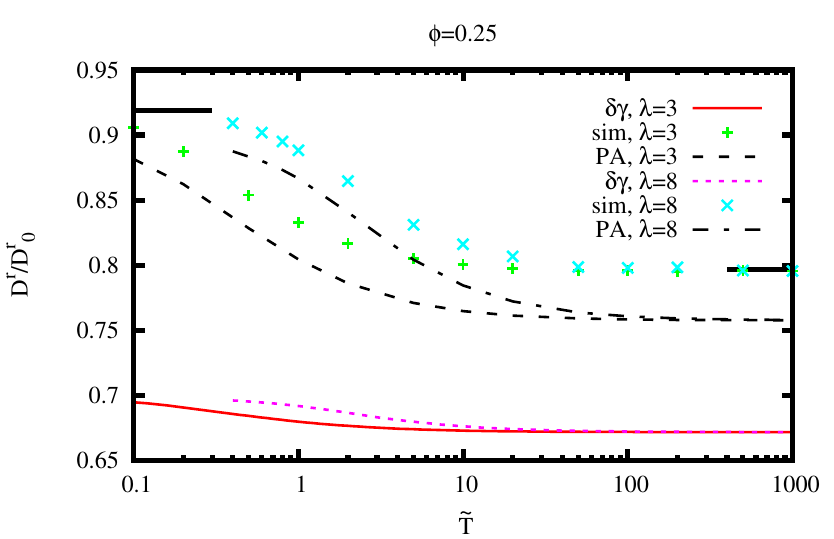}%
\vspace{1em}
\caption{Same as in Fig. \ref{fig dself rot 005} but for $\phi=0.25$.}%
\label{fig dself rot 025}%
\end{centering}
\end{figure}

\begin{figure}
\begin{centering}
\includegraphics[width=.6\textwidth]%
{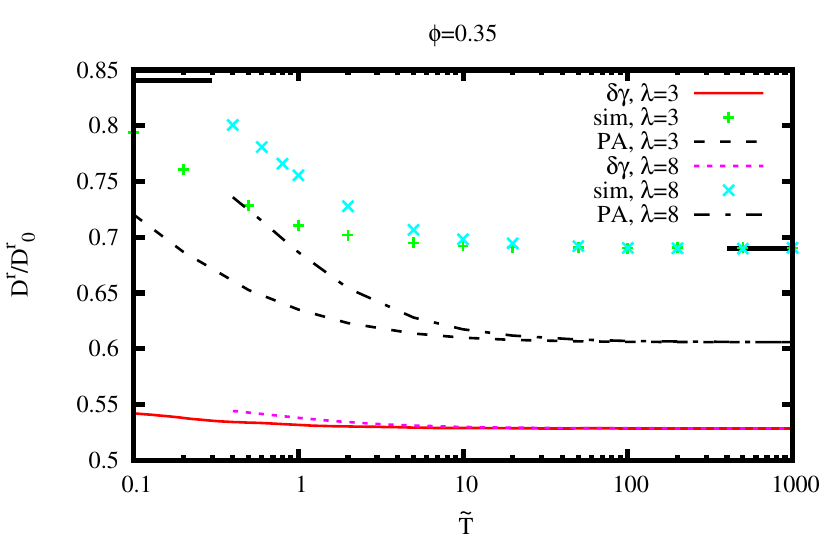}%
\vspace{1em}
\caption{Same as in Fig. \ref{fig dself rot 005} but for $\phi=0.35$.}%
\label{fig dself rot 035}%
\end{centering}
\end{figure}

Our simulation and theoretical results for $D^r/D_0^r$ as a function of the inverse reduced Yukawa interaction parameter  $\tilde{T}$ are depicted in Figs.~\ref{fig dself rot 005} - \ref{fig dself rot 035}, for volume fractions  $\phi=0.05 - 0.35$. In each figure, the results for the vertical fluid-phase pathway at $\lambda = 3$ directed towards the fluid-bcc phase coexistence line is compared with the results for the pathway at $\lambda = 8$ directed towards the fluid-fcc coexistence line (cf. Fig. \ref{fig phase diagram}). Note the different ordinate scales in the four figures selected 
to highlight the differences in the theoretical and simulation results for $D^r$. 

We start with the discussion of the simulation results for the HSY systems. The slowing influence of the HIs on rotational self-diffusion increases when the likelihood for near-distant particle pairs increases. According to our discussion of the RDFs in Fig. \ref{fig radial at contact}, 
the particles for $\lambda=8$ repel each other more strongly than those for $\lambda=3$, so that the radial region where $g(r)$ is small is more extended in the former case. This explains why all curves of $D^r$ for $\lambda=8$ are located above those for $\lambda=3$, for all considered volume fractions. The HSY particles can approach each other more closely with increasing $\tilde{T}$. This is reflected in curves for $D^r$ which are monotonically decreasing. 

As discussed earlier, at large values of $\tilde{T}$ a plateau region of $D^r$ is reached where the particles behave essentially as neutral hard spheres, independent of $\lambda$. The simulation curves in Figs.~\ref{fig dself rot 005}-\ref{fig dself rot 035} converge therefore for large $\tilde{T}$ towards the result in Eq. (\ref{eq:virial-HS-second})  which accurately describes the $\phi$-dependence of $D^r$ for neutral hard spheres up to the freezing transition volume fraction. With increasing $\phi$, the hard-sphere plateau region is reached for smaller values of $\tilde{T}$. In the opposite limit of low $\tilde{T}$ values, the interaction of the HSY particles is dominated by the Yukawa potential part. For smaller volume fractions where $g(\sigma^+) \approx 0$ is observed, Eq.~ (\ref{eq:CSdeionized}) derived originally for low-salinity charge-stabilized particles is expected to be a decent description of $D^r$ in the small-$\tilde{T}$ region. It is noticed that the simulation curves for $\lambda=3$ and $8$, and $\phi=0.05$ and $0.15$, are indeed converging, with decreasing $\tilde{T}$, towards the result in Eq. (\ref{eq:CSdeionized}). Differences are visible for $\phi=0.25$ and $0.35$ where Eq. (\ref{eq:CSdeionized}) provides only an upper bound for $D^r$. 

The simulation curves for both values of $\lambda$ are well reproduced by the PA approximation for $\phi=0.05$ and $0.15$. This is an expected feature of this method which becomes exact at low concentrations. At higher concentrations, three-body and higher-order HIs contributions come into play which are disregarded in the simple PA treatment. As a consequence, the simulation curves for $\phi=0.25$ and $0.35$ are underestimated at 
all values of $\tilde{T}$, i.e. the slowing influence of the HIs on rotational self-diffusion is overestimated. 
As it was noticed earlier in the context of translational self-diffusion \cite{heinen2010short}, this can be attributed to the fact that the PA approximation neglects shielding of the HIs between two particles by other particles in their vicinity. While $D^r$ at larger $\phi$ is underestimated in the PA approximation, 
the $\tilde{T}$ region where the simulation curves for $\lambda=3$ and $8$ converge is still well predicted. 
The PA method for $D^r$ is actually a decent approximation up to surprisingly large volume fractions. This should be contrasted with its performance for collective diffusion properties where its concentration range of application is significantly smaller \cite{heinen2011short}. \\               
We discuss now the performance of the revised second-order $\delta\gamma$ scheme.  
It is noticed from Figs.~\ref{fig dself rot 005} - \ref{fig dself rot 035} that it systematically, and significantly,  underestimates $D^r$, for all values of $\tilde{T}$ and all considered volume fractions. The relative error, $|D^r_{\delta\gamma}-D^r_{\textnormal{sim}}|/D^r_{\textnormal{sim}}$, increases systematically with increasing $\phi$, and it is more pronounced at the lower-$\tilde{T}$ side where the Yukawa repulsion is strong. In the hard-sphere-like interaction regime of large $\tilde{T}$ values, the relative error increases from about $1\%$ at $\phi=0.05$ to $23\%$ at $\phi=0.35$. In comparison, the relative error in the small $\tilde{T}$ region is larger, increasing from
about $2\%$ at $\phi=0.05$ to $32\%$ at $\phi=0.35$. The most significant feature of the (revised and non-revised) $\delta\gamma$ method at higher concentrations is its weak sensitivity to changes in range and strength of the pair potential, and to the accompanying changes in the RDF. For example, in Fig. \ref{fig dself rot 035} with $\phi=0.35$, the relative difference between the curves for $\lambda=3$ and $\lambda=8$ is four times larger for the simulation data than for the $\delta\gamma$ method curves. Incidentally, a similarly weak dependence on the shape of the pair potential and RDF has been found for the non-revised $\delta\gamma$ method result for the high-frequency viscosity of charge-stabilized suspensions \cite{heinen2011short}.

\begin{figure}
\begin{centering}
\includegraphics[width=.6\textwidth]%
{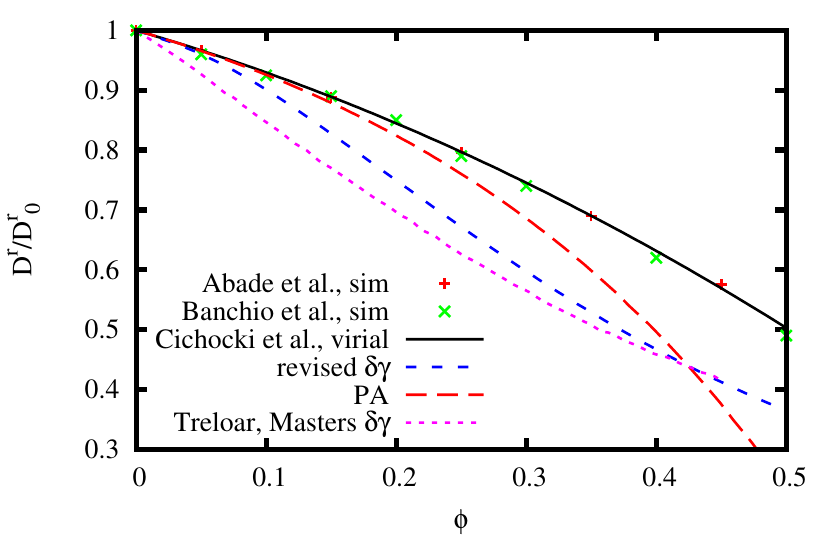}%
\vspace{1em}
\caption{Normalized rotational self-diffusion coefficient $D^{r}%
	/D_{0}^{r}$ of neutral no-slip hard spheres as a function of $\phi$. Simulation results by Abade {\em et al.} \cite{abade2011rotational} and Banchio {\em et al.} \cite{banchio2008short} are compared to the second-order virial expansion result in Eq.~(\ref{eq:virial-HS-second}) by Cichocki {\em et al.} \cite{trojczastkowasamodyfuzja}, 
the original $\delta\gamma$ method result by Treloar and Masters \cite{treloar1989short}, 
and our revised second-order $\delta\gamma$ method and  
PA predictions.
}
\label{fig dself hardspheres}%
\end{centering}
\end{figure}

The revised $\delta\gamma$ scheme for $D^r$ performs somewhat better for neutral hard spheres. A direct comparison with simulation results for hard spheres (where $\lambda=\infty$ or $\tilde{T}=\infty$) is made in Fig. \ref{fig dself hardspheres}. The simulation data are well described by Eq. (\ref{eq:virial-HS-second}) in the full liquid-phase concentration range . While the revised $\delta\gamma$ method for $D^r$ significantly improves the original second-order $\delta\gamma$ method results for hard spheres by Treloar and Masters in the range $\phi \leq 0.4$, there is no improvement at larger volume fractions. A general observation made for the HSY systems is that the relative mean difference between revised and non-revised second-order $\delta\gamma$ results for $D^r$ is typically $5-7\%$ or less.            

The low sensitivity of the $\delta \gamma$ scheme on the shape of the RDF can be  
related to its mean-field structure. 
An important ingredient of BM method 
is the effective propagator $G_{\left\langle \mathcal{M}_{R}\right\rangle }$, defined by Eq. (\ref{gmr}), 
which depends on $\phi$ but not on the RDF or higher-order static correlation functions.
The suspension microstructure for a given volume fraction enters into the BM approach only through 
fluctuations which are included up to second order.
The truncation of the renormalized fluctuation expansion at higher order than the second one could arguably enlarge 
the sensitivity of the method on the equilibrium suspension microstructure, for the price that triplet or even higher-order static distribution functions are then required as additional input.

Fig.~\ref{fig dself hardspheres} includes also the PA approximation prediction for hard spheres. The simulation data are well described by this method for $\phi \leq 0.2$, but $D^r$ is increasingly underestimated at larger $\phi$.  

\section{Summary and conclusions} \label{sec:conclusions}

We have presented the first comprehensive theory-simulation study of short-time rotational diffusion in suspensions in the fluid-like phase, with the particles interacting by the HSY potential. 
Since this effective pair potential is generic to many different soft matter systems including ionic microgels and globular protein solutions, the presented results should be of broad interest. 

A large body of high-precision simulation data was generated and compared with the results by two theoretical methods. The first and more elaborate one is a revised second-order version of the original Beenakker-Mazur method which has been   adapted to rotational diffusion by Treloar and Masters \cite{treloar1989short}, in the context of no-slip hard spheres. 
In our revised second-order $\delta\gamma$ method, various approximation steps made in the original method have been avoided, in particular regarding the treatment of the HIs which in the original method was rather approximate. The 
second method is the PA approximation with full account of two-body HIs contributions including lubrication terms, but with three-body and higher-order HIs contributions disregarded. 

General features of $D^r$ observed in our simulation study are its monotonic decrease with increasing $\tilde{T}$, reproduced qualitatively by both theoretical methods, and its strong sensitivity on the Yukawa potential range parameter $\lambda$ for intermediate values of $\tilde{T}$. This sensitivity is well captured by the PA method, different from the revised $\delta\gamma$ method which  shows this sensitivity 
for low concentrations only. A lower bound of $D^r$ is provided by the Eq.~(\ref{eq:virial-HS-second}) for hard spheres, reached by the simulation curves of $D^r$ at large values of $\tilde{T}$. An upper bound is given by the scaling result in Eq.~(\ref{eq:CSdeionized}). This bound is approached by the simulation curves at low values of $\tilde{T}$, provided $\phi$ is sufficiently small (i.e., $\phi \lesssim 0.15$)  
and $\lambda$ not very large (i.e., $\lambda \lesssim 8$). 

Even though HIs are accounted for to significantly higher accuracy than in BM and Treloar and Masters approach, the resulting improvement of $D^r$ in our revised second-order $\delta\gamma$ method is comparatively small, amounting roughly to $5 - 7\%$ for $\phi \lesssim 0.4$. A similar observation has been made regarding the hydrodynamic function, the short-time translational self-diffusion coefficient, and the high-frequency viscosity of hard spheres \cite{makuch2012transport}. This may be due to the interplay of mean-field and HIs approximations going into the $\delta\gamma$ method which can cause uncontrolled fortuitous cancellations or fortifications of errors.             

The (revised) second-order $\delta\gamma$ method performs distinctly better for collective than 
self-diffusion properties, namely for the wavenumber dependent distinct part of the hydrodynamic function and the collective diffusion coefficient \cite{heinen2011short}. In particular regarding the distinct hydrodynamic function part, the $\delta\gamma$ method performs quite well both for hard spheres and charge-stabilized particles. 
In future work, it will be interesting to find out whether the overall good agreement with simulation data for the collective diffusion properties of HSY systems can be further improved by the revised method.  
The performance of the revised $\delta\gamma$ method regarding $D^r$ can be possibly improved by 
the inclusion of third-order renormalized fluctuation contributions where, however, static three-body distribution functions are required as input to the extended method in addition to $g(r)$. 

The PA method with its full account of two-body HIs contributions describes the HSY simulation data for $D^r$ quite well for volume fractions up to $\phi \approx 0.2$. At larger concentrations, however, 
the rotational self-diffusion coefficient is underestimated 
owing to the disregarded hydrodynamic shielding effect 
embodied in three-body and higher-order hydrodynamic mobility tensor contributions.     
In going beyond the PA approximation, three-body irreducible hydrodynamic cluster contributions could be 
additionally considered. For low-salinity systems, the leading-order far-distant three-body contributions have been accounted for in \cite{zhang2002tracer}, in conjunction with Kirkwood's superposition approximation for the static three-body distribution function. The long-distance three-body cluster contribution to $D^r$ is positive valued, with the effect of bringing the value for $D^r$ thus closer to the simulation data \cite{zhang2002tracer, koenderink2003validity}. 

Finally, we note that both the revised $\delta\gamma$ method and the PA approach can be rather straightforwardly extended to colloidal particles with internal hydrodynamic structure, and hydrodynamic surface boundary conditions different from the no-slip one used in the present work. This offers the possibility to study theoretically, e.g., the rotational self-diffusion of weakly crosslinked ionic and non-ionic microgels, and of core-shell particles with a fluid-permeable soft shell. 

\section*{Acknowledgements}

K.M. has been supported by MNiSW grant IP2012 041572, and, at the earlier stage of the research,
also acknowledged support by the Foundation for Polish Science (FNP) through the TEAM/2010-6/2 project, co-financed by the EU European Regional Development Fund.
M.H. acknowledges support by a fellowship within the Postdoc-Program of 
the German Academic Exchange Service (DAAD).
G.C.A. acknowledges financial support from CNPq (480018/2013-8) and
expresses his deep gratitude to Prof. Eligiusz Wajnryb for making his
HYDROMULTIPOLE code available for this work. Numerical HYDROMULTIPOLE
calculations were performed at NACAD-COPPE/UFRJ in Rio de Janeiro,
Brazil.

\appendix

\section*{Appendix: Hydrodynamic matrices}

The translational and rotational mobility matrices with tensor elements
$\boldsymbol{\mu}_{ij}^{rr}\left( \mathbf{R}_{1}\ldots\mathbf{R}_{N}\right)$
and 
$\boldsymbol{\mu}_{ij}^{tt}\left( \mathbf{R}_{1}\ldots\mathbf{R}_{N}\right)$
appearing in Eqs. \eqref{rot self diff} and \eqref{eq: tran self diff} for the short-time rotational and translational 
self-diffusion coefficients, respectively, can be calculated using the hydrodynamic multipole matrices
$Z_0$, $\mu_0$, $\hat{Z}_0$, and $G$ introduced in Ref. \cite{cichocki2000friction,Cichocki2002three}.
Each of these matrices is indexed by the set of three indices $\left(l,m,\sigma\right)$ with 
$l=1,\ldots,\infty$, $m=-l,-l+1,\ldots,l$, and $\sigma=0,1,2$.
The matrix $Z_0$ is given by the formula
\begin{equation}
	[ Z_0(\mathbf{R}_i) ]_{l m \sigma, l' m' \sigma'}=
	\delta_{ll'}\delta_{mm'}\eta (2a)^{2l+\sigma+\sigma'-1}z_{l,\sigma \sigma'}\,,
\end{equation}
where the dimensionless coefficients $z_{l,\sigma \sigma'}$ have been defined in Ref. \cite{Cichocki2002three}.
The only non-vanishing elements of the matrix $\mu_0$ are for
$l=l'=1$ and $\sigma=\sigma'=0$, with
\begin{equation}
	[\mu_0(\mathbf{R}_i)]_{l m 0, l' m' 0}=
	\delta_{ll'}\delta_{mm'}\frac{2}{9\eta a}\,,
\end{equation}
and for $l=l'=1$ and $\sigma=\sigma'=1$, with
\begin{equation}
	[\mu_0(\mathbf{R}_i)]_{l m 0, l' m' 0}=
	\delta_{ll'}\delta_{mm'}\frac{1}{6\eta a^3}\,.
\end{equation}
The matrix $\hat{Z}_0$ is related to $\mu_0$ and $Z_0$ by
\begin{equation}
	\hat{Z}_0(\mathbf{R}_i)=Z_0(\mathbf{R}_i)-Z_0(\mathbf{R}_i)\mu_0(\mathbf{R}_i)Z_0(\mathbf{R}_i)\,.
\end{equation}
The Oseen tensor in multipole space, $G(\mathbf{R}_i,\mathbf{R}_j)$,
is for non-overlapping configurations given by
\begin{equation}
	[G(\mathbf{R}_i,\mathbf{R}_j)]_{l m \sigma, l' m' \sigma'}=
	\frac{n_{lm}}{\eta n_{l'm'}}S^{+-}(\mathbf{R}_i-\mathbf{R}_j,lm\sigma,l'm'\sigma')
	\ \ \ \ \textnormal{for }|\mathbf{R}_i-\mathbf{R}_j|>2a \,,
\end{equation}
where the coefficients $S^{+-}$ and $n_{lm}$ have been introduced in Ref. \cite{felderhof1989displacement}.
The matrices noted above are used to construct the scattering series for the generalized multipole mobility matrix 
elements  
according to 
\begin{eqnarray}
\mu_{ij}\left( \mathbf{R}_{1}\ldots\mathbf{R}_{N}\right) & = &
\delta_{ij}\mu_0\left( \mathbf{R}_{i}\right) 
+ \left(  1-\delta_{ij}\right) 
\mu_0\left(  \mathbf{R}_{i}\right) Z_0\left(  \mathbf{R}_{i}\right)
G\left(  \mathbf{R}_{i},\mathbf{R}_{j}\right)
Z_0\left( \mathbf{R}_{j}\right)  \mu_0\left( \mathbf{R}_{j}\right)  + \nonumber \\
&&+\sum_{\substack{k=1,\\k\neq i,k\neq j}}^{N}
\mu_0\left(  \mathbf{R}_{i}\right) Z_0\left(  \mathbf{R}_{i}\right)
G\left(  \mathbf{R}_{i},\mathbf{R}_{k}\right) 
\hat{Z}_0\left( \mathbf{R}_{k}\right) 
G\left(  \mathbf{R}_{k},\mathbf{R}_{j}\right) 
Z_0\left( \mathbf{R}_{j}\right)  \mu_0\left( \mathbf{R}_{j}\right) 
  \nonumber \\
  && +\ldots \,\,.	\label{sca ser gen mob mat}
\end{eqnarray}

In the multipole basis, the lowest multipole elements $l=1,\sigma=0$ correspond to translational motion.
Therefore, the tensorial elements $\boldsymbol{\mu}_{ij}^{tt}$ of the translational mobility matrix can be calculated from the 
generalized mobility matrix with elements $\mu_{ij}$ by an appropriate projection.
To this end, the projector $\overline{P}^t$ is introduced by
\begin{equation}
	[\overline{P}^t]_{\alpha,lm\sigma} = \delta_{l1} 
	\delta_{\sigma 0} \sqrt{\frac{3}{4 \pi}} [\mathbf{y}_{1m}]_{\alpha}\,,
\end{equation}
where $\alpha=1,2,3$ denote three Cartesian components. The tensor $\mathbf{y}_{1m}=X\mathbf{y}^{(R)}_{1m}$ is defined in Ref. \cite{marysiaElekFragment}, together with $X$ and $\mathbf{y}^{(R)}$. Moreover,  
$\overline{P}^{t \dagger}$ is the Hermitean conjugate of $\overline{P}^t$.
Finally, the Cartesian mobility matrix is related to the generalized multipole mobility matrix by the projection operation
\begin{equation}
\boldsymbol{\mu}_{ij}^{tt}\left( \mathbf{R}_{1}\ldots\mathbf{R}_{N}\right)
= \overline{P}^t \mu_{ij}\left( \mathbf{R}_{1}\ldots\mathbf{R}_{N}\right) \overline{P}^{t \dagger} \,.
\end{equation}
The according expression for the rotational mobility matrix reads
\begin{equation}
\boldsymbol{\mu}_{ij}^{rr}\left( \mathbf{R}_{1}\ldots\mathbf{R}_{N}\right)
= \overline{P}^r \mu_{ij}\left( \mathbf{R}_{1}\ldots\mathbf{R}_{N}\right) \overline{P}^{r \dagger} \,.
\label{app rot mob mat}
\end{equation}
Here, the operator $\overline{P}^r$ projects from the multipole space the elements $l=1,\sigma=1$ to the Cartesian components corresponding to rotational motion, i.e.
\begin{equation}
	[\overline{P}^r]_{\alpha,lm\sigma} = 
	\delta_{l1} \delta_{\sigma 1} \sqrt{\frac{3}{4 \pi}} [\mathbf{y}_{1m}]_{\alpha}\,.
\end{equation}

In the scattering series for the mobility matrix in Eq. \eqref{sca ser gen mob mat}, 
the single particle matrices $\mu_0 Z_0$ and $Z_0 \mu_0$ at the start and end of a scattering sequence
are different from the single-particle matrices $\hat{Z}_0$ dispersed in between the propagators $G$.
In the derivation of the renormalized fluctuation expansion, it is thus very useful to rewrite
the scattering series in a way that all the single particles matrices
starting a scattering sequence or appearing in between the propagators $G$ are the same.

This can be achieved by introducing the matrix defined by \cite{makuch2012scattering}
\begin{equation}
M=\left [ 
\begin{array}{cc}
	\mu_0  & \mu_0 Z_0 \\
       Z_0 \mu_0 & \hat{Z}_0 
\end{array}
\right ]\,. 
\end{equation}
According to the above definition, the matrix $M$ has in addition to the multipole indexes $l,m,\sigma$ the index $u=1,2$.
Therefore, $M_{1lm\sigma,1l'm'\sigma'}=[\mu_0]_{lm\sigma,l'm'\sigma'}$, $M_{1lm\sigma,2l'm'\sigma'}=[\mu_0 Z_0]_{lm\sigma,l'm'\sigma'}$ et cetera.  
Similarly, we define the matrix $[G]_{ulm\sigma,u'l'm'\sigma'}$ by
\begin{equation}
G=\left [ 
\begin{array}{cc}
	0  & 0 \\
       0 & G 
\end{array}
\right ]\,, 
\end{equation}
and generalize the projectors $\overline{P}^t$ and $\overline{P}^r$ to
\begin{equation}
	[P^t]_{\alpha,ulm\sigma} = \delta_{u1} \delta_{l1} \delta_{\sigma 0} \sqrt{\frac{3}{4 \pi}} [\mathbf{y}_{1m}]_{\alpha}\,,
\end{equation}
\begin{equation}
	[P^r]_{\alpha,ulm\sigma} = \delta_{u1} \delta_{l1} \delta_{\sigma 1} \sqrt{\frac{3}{4 \pi}} [\mathbf{y}_{1m}]_{\alpha}\,.
\end{equation}
In the extended space expressions in Eqs.~\eqref{sca ser gen mob mat} and \eqref{app rot mob mat}, 
the rotational mobility matrix can be rewritten in the form given by Eq. \eqref{mu rr}.
This can be done analogously for the translational mobility matrix, 
with the only difference lying in the different projection operators.

\bibliographystyle{aipnum4-1}
\bibliography{baza_artukolow}

~\\~\\~\\

\end{document}